\documentclass[12pt,epsf,amstex]{article}
\usepackage [dvips]{graphicx}
\usepackage{amsmath}
\usepackage{amssymb}
\usepackage{epsfig}
\usepackage{verbatim}

\addtocounter{secnumdepth}{1}
\font\capfont=cmbx12 at 50 pt 
\newbox\capbox \newcount\capl \def\a{A}
\def\docappar{\medbreak\noindent\setbox\capbox\hbox{%
\capfont\a\hskip0.15em}\hangindent=\wd\capbox%
\capl=\ht\capbox\divide\capl by\baselineskip\advance\capl by1%
\hangafter=-\capl%
\hbox{\vbox to8pt{\hbox to0pt{\hss\box\capbox}\vss}}}
\def\cappar{\afterassignment\docappar\noexpand\let\a }

\begin{document}
\newcommand{\barr}{\bar{r}}
\newcommand{\bbeta}{\bar{\beta}}
\newcommand{\bgamma}{\bar{\gamma}}
\newcommand{\half}{\frac{1}{2}}
\newcommand{\summ}{\sum_{m=1}^n}
\newcommand{\tsum}{\Sigma}
\newcommand{\sumqno}{\sum_{q\neq 0}}
\newcommand{\bsA}{\mathbf{A}}
\newcommand{\bsV}{\mathbf{V}}
\newcommand{\bsE}{\mathbf{E}}
\newcommand{\bsZ}{\hat{\mathbf{Z}}}
\newcommand{\bse}{\mbox{\bf{1}}}
\newcommand{\bspsi}{\hat{\boldsymbol{\psi}}}
\newcommand{\cdottt}{\!\cdot\!}
\newcommand{\deltaR}{\delta\mspace{-1.5mu}R}

\newcommand{\bGamma}{\boldmath$\Gamma$\unboldmath}
\newcommand{\dd}{\mbox{d}}
\newcommand{\ee}{\mbox{e}}
\newcommand{\p}{\partial}

\newcommand{\bA}{\mathbf{A}}
\newcommand{\bB}{\mathbf{B}}
\newcommand{\bc}{\mathbf{c}}
\newcommand{\bC}{\mathbf{C}}
\newcommand{\bD}{\mathbf{D}}
\newcommand{\bF}{\mathbf{F}}
\newcommand{\ff}{\mathbf{f}}
\newcommand{\bL}{\mathbf{L}}
\newcommand{\tell}{\tilde{\ell}}
\newcommand{\bM}{\mathbf{M}}
\newcommand{\bO}{\mathbf{O}}
\newcommand{\bq}{\mathbf{q}}
\newcommand{\br}{\mathbf{r}}
\newcommand{\rav}{r_{\rm av}}
\newcommand{\ravc}{r_{\rm av,c}}
\newcommand{\bR}{\mathbf{R}}
\newcommand{\btR}{\mathbf{\tilde{R}}}
\newcommand{\tR}{\tilde{R}}
\newcommand{\bs}{\mathbf{s}}
\newcommand{\bS}{\mathbf{S}}
\newcommand{\sav}{{\sf{s}}}
\newcommand{\tav}{t_{\rm av}}
\newcommand{\bT}{\mathbf{T}}
\newcommand{\bv}{\mathbf{v}}
\newcommand{\bx}{\mathbf{x}}

\newcommand{\cA}{{\cal{A}}}
\newcommand{\cV}{{\cal{V}}}
\newcommand{\cK}{{\cal{K}}}
\newcommand{\cQ}{{\cal{Q}}}
\newcommand{\pcV}{{\pmb{\cal{V}}}}
\newcommand{\Vsp}{V^{\rm sp}}
\newcommand{\Aop}{Y_{\Lambda}}
\newcommand{\Xop}{F_4}
\newcommand{\Uop}{G_2}
\newcommand{\Gtwo}{G_2}
\newcommand{\tFfour}{\tilde{F}_4}
\newcommand{\Ffour}{F_4}
\newcommand{\tFtwo}{\tilde{F}_2}
\newcommand{\Ftwo}{F_2}
\newcommand{\Atwo}{A_2}
\newcommand{\Afour}{A_4}
\newcommand{\kap}{\kappa}

\newcommand{\sfA}{{\sf A}}
\newcommand{\sfB}{{\sf B}}
\newcommand{\sfC}{{\sf C}}
\newcommand{\sfL}{{\sf L}}
\newcommand{\sfR}{{\sf R}}
\newcommand{\sfS}{{\sf S}}

\newcommand{\la}{\langle}
\newcommand{\ra}{\rangle}
\newcommand{\ranul}{\rangle_{_{\!0}}}
\newcommand{\bigranul}{{\big\rangle}_{_{\!0} }}
\newcommand{\raone}{\rangle_{_1}}
\newcommand{\rao}{\rangle\raisebox{-.5ex}{$\!{}_0$}}  
\newcommand{\rae}{\rangle\raisebox{-.5ex}{$\!{}_1$}}
\newcommand{\raG}{\rangle_{_{\!G}}}
\newcommand{\rainr}{\rangle_r^{\rm in}}
\newcommand{\beq}{\begin{equation}}
\newcommand{\eeq}{\end{equation}}
\newcommand{\bea}{\begin{eqnarray}}
\newcommand{\eea}{\end{eqnarray}}
\newcommand{\non}{\nonumber\\[2mm]}
\def\lsim{\:\raisebox{-0.5ex}{$\stackrel{\textstyle<}{\sim}$}\:}
\def\gsim{\:\raisebox{-0.5ex}{$\stackrel{\textstyle>}{\sim}$}\:}

\numberwithin{equation}{section}

\thispagestyle{empty}
\title{{\Large {\bf Exact asymptotic statistics\\[3mm] 
of the $n$-edged face\\[3mm] 
in a 3D Poisson-Voronoi tessellation\\
\phantom{xxx} }}}
 
\author{{H.\,J. Hilhorst}\\[5mm]
{\small Laboratoire de Physique Th\'eorique, B\^atiment 210}\\ 
{\small Universit\'e Paris-Sud and CNRS}\\[-1mm]
{\small Universit\'e Paris-Saclay}\\[-1mm]
{\small 91405 Orsay Cedex, France}\\}

\maketitle
\begin{small}
\begin{abstract}
\noindent 
We consider the 3D Poisson-Voronoi tessellation.
We investigate the joint probability distribution
$\pi_n(L)$ for an arbitrarily selected cell face to be $n$-edged
and for the distance between the seeds of its adjacent cells to
be equal to $2L$.
We derive an exact expression for this quantity, valid in the limit
$n\to\infty$ with $n^{1/6}L$ fixed.
The leading order correction term is determined.
Good agreement with earlier Monte Carlo data is obtained.
The cell face is surrounded by a three-dimensional excluded domain that is the
union of $n$ balls; it is pumpkin-shaped and
analogous to the flower of the 2D 
Voronoi cell. For $n\to\infty$ this domain tends towards a torus 
of equal major and minor radii. The radii scale as $n^{1/3}$, 
in agreement with earlier heuristic work.
We achieve a detailed understanding
of several other statistical properties of the $n$-edged cell face.
\\

\end{abstract}
\vspace{10mm}

\noindent LPT -- ORSAY 16/01\\
{\small $^1$Laboratoire associ\'e au Centre National de la
Recherche Scientifique - UMR 8627}

\thispagestyle{empty}
\newpage


\section{Introduction}
\label{secintroduction}

\cappar Let there be given a set of $N$ point-like ``seeds'' in a domain of
volume $V$ in three-dimensional Euclidean space $\mathbb{R}^3$. 
The Voronoi tessellation based on this set is the partitioning 
of space into cells, one around each seed, in such a way that every 
generic point of space is in the cell of the seed to which it is closest. 
If the seeds are randomly and uniformly distributed,
the tessellation is called a Poisson-Voronoi tessellation.
Obviously this construction is easily generalized to arbitrary spatial 
dimension.

Voronoi tessellations 
have applications across the sciences,
whether as models that directly describe natural systems 
or as tools for data analysis. 
Many applications have been reviewed by Okabe {\it et al.} \cite{Okabeetal00}.

The exact statistics of Poisson-Voronoi cells has been a subject of 
investigation by physicists and mathematicians alike.
Lists of exact results are given in Ref.\,\cite{Okabeetal00}
for tessellations of $\mathbb{R}^2$ and $\mathbb{R}^3$.
They refer to properties in the ``thermodynamic limit'',
that is, the limit $N,V\to\infty$ at fixed seed density $\lambda=N/V$.
In this paper we derive new exact results in 3D.
They concern the statistics of a face shared by two 3D cells,
and in particular the limit in which the number of edges of that face
becomes very large. The work builds on earlier results
in 2D that we briefly summarize in the next subsection.


\subsection{The 2D Poisson-Voronoi tessellation}
\label{sec2D}

In two-dimensional space a quantity of basic interest is the 
sidedness distribution $p_n^{(2)}$, that is, the probability 
for a two-dimensional Poisson-Voronoi cell to have exactly $n$ sides. 
No simple exact closed-form expression is known for this
elementary probability distribution.

About a decade ago we investigated \cite{Hilhorst05a,Hilhorst05b}
the large $n$ behavior of $p_n^{(2)}$ and found the 
asymptotic expansion of $\log p_n^{(2)}$ in inverse powers of $n$. 
One by-product of this calculation was
an efficient algorithm \cite{Hilhorst07}
for simulating $n$-sided cells for large $n$.
Another development \cite{Hilhorst06} based on Ref.\,\cite{Hilhorst05b}
dealt with the relation between a many-sided cell and its first-neighbor 
cells and led to the replacement of Aboav's ``linear law''
by a square-root law.

The methods of Refs.\,\cite{Hilhorst05a,Hilhorst05b} proved to be 
applicable to at least two other two-dimensional geometric problems. 
The first application \cite{HilhorstCalka08} is to a family of line tessellations
introduced by Hug and Schneider \cite{HugSchneider07}.
The second application \cite{HilhorstCalkaSchehr08} is to
``Sylvester's question''\cite{Sylvester1864}: 
if $n$ random points are distributed uniformly
in some convex subdomain of $\mathbb{R}^2$,
then what is the probability $p^{\rm Sylv}_n$ that 
they are the vertices of a convex $n$-gon?
For the subdomain
equal to the unit disk, Ref.\,\cite{HilhorstCalkaSchehr08}  obtained 
the asymptotic expansion of $\log p^{\rm Sylv}_n$.

In all these problems there appears a closed random curve 
that in the limit $n\to\infty$ tends to a circle while satisfying a stochastic ordinary linear second order differential equation known as the {\it random acceleration process\,}
\cite{Burkhardt07,Majumdaretal10,Reymbautetal11}. 
These interrelationships provide a motivation for 
the study of this paper, in which for the first time our methods are brought
to bear on a 3D question.


\subsection{The edgedness of a face between 3D cells}
\label{secedgedness}

One immediate 3D generalization of the 2D sidedness distribution
is the facedness probability $p_n^{(3)}$, that is, the
probability that a three-dimensional Poisson-Voronoi cell have $n$ faces. 
Again, 
this probability distribution is unknown and one might hope to find its asymptotic large-$n$ behavior by the methods of 
Refs.\,\cite{Hilhorst05b,HilhorstCalka08,HilhorstCalkaSchehr08}. 
However, we do not know how to solve that problem.

A different generalization of the 
2D quantity $p^{(2)}_n$ to 3D is the probability 
-- henceforth to be denoted for simplicity by $p_n$ --
that an arbitrarily chosen face shared by two 3D cells have exactly $n$ edges.
In this paper we will find 
the asymptotic expansion of $\log p_n$ for large $n$
together with a large number of other statistical properties,
summarized below. 

This question about the edgedness distribution is richer than the one about the facedness.
The facedness question involves a single 3D Voronoi cell
and is, statistically, spherically symmetric.
The edgedness question, however, 
involves two adjacent 3D cells (the ``focal cells'') and only has
rotational symmetry about the axis passing through 
the seeds of the two cells.
The distance $2L$ between these seeds (where $L$ is called the ``focal distance'') enters the game as an extra parameter.
\vspace{3mm}

We will be naturally led to consider the joint probability distribution 
$\pi_n(L)\dd L$, defined as the probability that an arbitrarily chosen cell face
have $n$ edges and is shared by two cells 
whose focal distance is between $L$ and $L+\dd L$.
As a consequence 
\beq
p_n = \int_0^\infty\!\dd L\,\pi_n(L).
\label{intropinL}
\eeq
We will also write
\beq
\pi_n(L) = Q_n(L)p_n\,,
\label{introQnL}
\eeq
where $Q_n(L)$ is the conditional probability that an $n$-edged cell face
separate seeds of distance $2L$.
The objects of interest in this paper are $p_n$ and
$Q_n(L)$, both in the limit of large $n$. 
\vspace{3mm}

Earlier studies of the edgedness were performed by 
Kumar {\it et al.} \cite{Kumaretal92} and more recently 
in Ref.\,\cite{HilhorstLazar14}, where
the Monte Carlo work of Lazar {\it et al.} \cite{Lazaretal13} was extended
and a heuristic theory was presented to explain the results.
In sections \ref{seccomparisonHL} and \ref{seccomparisonMC} we will compare
our present results to these earlier studies. 


\subsection{Method}
\label{secmethod}

In section \ref{secnandL} the edgedness probability $\pi_n(L)$ is cast 
in the form of a phase space
integral on the $3n$ position coordinates $\bR_1,\bR_2,\ldots,\bR_n$ in $\mathbb{R}^3$ of the first-neighbor seeds.
This multiple integral is then analogous to the configurational partition function of a system of $n$ interacting particles.
The following sections are basically a concatenation of steps needed to
evaluate this integral.

In section \ref{secintTheta} we perform a 
transformation to new radial and polar coordinates.
In section \ref{secexcludedvolume} 
we establish the shape of a 3D excluded domain which is analogous to the
``flower'' of the 2D Voronoi cell. There are good reasons in our case to call
this domain a ``pumpkin.''
We point out by geometrical considerations 
that there is an invariance allowing
this 3D problem to be reduced to one in 2D. We do precisely that
in section \ref{secintegratingtheta} by 
integrating over the polar angles.
The techniques developed in  Ref.\,\cite{Hilhorst05b} may then
be adapted to the resulting 2D problem.
In section \ref{secrewriting2n} we perform further 
coordinate transformations that leave it as a problem of integrating on a single radial and $2n-1$ angular variables.
All these rewritings of the original problem are reversible
and merely amount to a different representation of $\pi_n(L)$.
\vspace{3mm}

In section \ref{seclargenexpansion} we prepare for a large-$n$ expansion of
$\pi_n(L)$.
Our strategy is to hypothesize, in subsection \ref{secscalingwithn},
the appropriate scaling with $n$ of all variables of integration involved, 
and to show that these assumed scalings are consistent and lead to an expansion with finite coefficients.
In subsections \ref{secexpansioncV}-\ref{secexpansionKcV} we carry out the expansion
of the various factors
in the integrand in negative powers of $n$ to the order required.
In section \ref{secintegratingsav} we perform the radial integration by means of a saddle point calculation. In sections \ref{sectransforming}, \ref{seccalculationexpV}, and \ref{seccalculationcR} 
we turn to the remaining integrals, 
which are those over the angular variables.
Their calculation is analogous to the 2D problem \cite{Hilhorst05b} and
we omit details. 
The work nevertheless goes beyond a simple analogy
in two respects. First, the present problem has the extra parameter $L$; and
second, our calculation of $Q_n(L)$ includes the next-to-leading 
terms, which are of relative order $n^{-1}$.
It will appear that these correction terms
greatly enhance the agreement with the simulations.

In section \ref{secarriving}
we combine all preceding relations to arrive at the final results.
In section \ref{secconclusion} we briefly conclude.
\vspace{3mm}

The calculation of this paper is of considerable length, even with our
omitting details that may be found in earlier papers.
However, only standard methods of mathematical analysis are used.

Since the subject matter of this work has attracted much activity among
mathematicians, the following remark may be appropriate.
We present our methods and results as ``exact,'' and they certainly are 
by the usual standards of theoretical physics. 
They are based on a formal expansion without, however, the necessary
proofs that the higher order terms are actually negligible for $n\to\infty$.
We are well aware that this procedure is not rigorous 
and fully accept that mathematicians consider our results as conjectures.


\subsection{Results}
\label{secresults}

The main results of this work are the following.

(i) In the large-$n$ limit the $n$-edged face tends towards a circle whose
random radius is narrowly peaked around
\beq
\sfR_n \simeq (2\pi^2\lambda)^{-1/3}n^{1/3}, \qquad n\to\infty.
\label{resultRn}
\eeq
Here and throughout the symbol $\simeq$ will indicate asymptotic equality.

(ii) The leading order behavior of $p_n$ is
\beq
p_n = \frac{1}{c_F}\left(\frac{3}{2\pi^5 n}\right)^{1/2}
\frac{(12\pi^2)^n}{(2n)!}\,C(3)\,
\left[ 1\,+\,{\cal O}(n^{-1}) \right]
\label{resultpn}
\eeq
in which $c_F=48\pi^2/35+2=15.535$ is the average number of faces of a cell
\cite{Okabeetal00} and $C(3)$ is a constant given by
\beq
C(3) = \prod_{q=1}^\infty \frac{q^4}{q^4+9} =  0.053891. 
\label{resC3}
\eeq
This constant may be interpreted as the partition function of the 
elastic degrees of freedom of the face,%
\footnote{{\it I.e.,} the deviations of the face boundary from
  circularity. This ``elasticity'' is of course of purely entropic origin.}
each factor in the product representing the contribution of a Fourier mode
of a definite wavenumber. A similar 
constant was found in our study 
\cite{Hilhorst05b} of the sidedness probability $p^{(2)}_n$ of the 2D cell.
\vspace{2mm}

(iii) The conditional
probability distribution $Q_n(L)$ defined in (\ref{introQnL})
may be expressed with the aid of
the scaling variable
\beq
y\equiv \kap n^{1/6}L, \qquad \kap = 2^{-1/6}3^{-1/2}\pi^{7/6}\lambda^{1/3}.
\label{xyc}
\eeq
In the limit $n\to\infty$ at fixed $y$ we have
\beq
Q_n(L) = \kap n^{1/6}\,\cQ(y)\left[1  + \frac{q_2 y^2+q_4 y^4-q_0}{n} 
+ {\cal O}(n^{-2})\right].
\label{resultQ}
\eeq
with 
\beq
\cQ(y)=\frac{32}{\pi^2}\,\exp\left( -\frac{4y^2}{\pi} \right), \qquad y>0.
\label{resultQy}
\eeq
The coefficients $q_i$ in (\ref{resultQ})
have explicit analytic expressions, given in section \ref{secarriving},
whose numerical values are
$q_2 = 3.71726,\, q_4 = 0.36025$, and $q_0 = 5.21263$.
In section \ref{seccomparisonMC} we compare (\ref{resultQ}) to the Monte Carlo
results of Ref.\,\cite{HilhorstLazar14} and find good qualitative agreement.
\vspace{2mm}

(iv)
It follows that for an $n$-edged face
the average value of $L$, to be denoted by $\sfL_n$, is given by
\beq
\sfL_n = \frac{1}{\kap n^{1/6}}
\Big[ 1+ \frac{l_1}{n} + {\cal O}(n^{-2}) \Big]
\label{resultLn}
\eeq
with an analytic expression for $l_1$ given in section \ref{secarriving} 
whose numerical value is $l_1=1.95558$.
The proportionality to $n^{-1/6}$ that appears in (\ref{resultLn}) may be seen as an attraction of entropic
origin between the two focal seeds. 
Comparison of (\ref{resultLn}) to the Monte Carlo results of Ref.\,\cite{HilhorstLazar14}
shows again good agreement.
\vspace{2mm}

(v) We denote as {\it first-neighbor seeds\,} those whose cells share an edge
of the face between the two focal cells.
The first-neighbor seeds lie in a shell whose width tends to zero for $n\to\infty$ and whose shape tends to the surface of a spindle torus of 
major radius $\sfR_n$ and minor radius 
$\sfS_n=\sqrt{\sfR_n^2-\sfL_n^2}$. In the limit $n\to\infty$ 
this spindle torus becomes a torus
with equal major and minor radius ( a ``horn torus'').


\section{The $n$-edged face and the focal distance $L$}
\label{secnandL}

\subsection{The joint probability $\pi_n(L)$}
\label{secdpinL}

We consider a Poisson-Voronoi tessellation in three-dimensional space,
constructed from $N$ seeds having
positions $\bR_1,\bR_2,\ldots,\bR_N$ that are independently and
uniformly distributed in a domain of volume $V$.
At appropriate points in the calculation we will let 
$N,V\to\infty$ with the seed density $\lambda=N/V$ fixed.
We may scale $\lambda$  to unity but will keep it in the formulas
a dimensional check.

We select an arbitrary cell face.
It is known \cite{Okabeetal00} that for $N,V\to\infty$
a cell has on average
\beq
c_F = \frac{48\pi^2}{35}+2 = 15.535
\label{xcF}
\eeq
faces. 
Since each face belongs to a unique pair of neighboring cells, 
selecting a cell face uniformly among all $\tfrac{1}{2}Nc_F$ faces amounts to
selecting a cell pair $(i,j)$ uniformly among all $\tfrac{1}{2}N(N-1)$ cell pairs
and retaining it only if $i$ and $j$ are neighbors.
The probability $P$ for retention is therefore 
\beq
P \simeq \frac{c_F}{N}\,, \qquad N\to\infty.
\label{cp}
\eeq
We may decompose $P$ according to
\beq
P = \int_0^\infty\dd L\,\sum_{n=3}^\infty P_n(L),
\label{dpnL}
\eeq
in which $P_n(L)\dd L$ is the probability that $i$ and $j$ be neighbors,
that the face they share be $n$-edged, 
and that their focal distance (half the distance between their seeds)
be between $L$ and $L+\dd L$. Hence
\beq
\pi_n(L)\dd L =\frac{P_n(L)}{P}\dd L
\label{dpinL}
\eeq
is the probability that an arbitrarily selected cell face be $n$-edged and
that the two cells sharing it have a focal distance between 
$L$ and $L+\dd L$. The normalization is
\beq
\int_0^\infty \dd L \sum_{n=3}^\infty \pi_n(L) = 1.
\label{normpinL}
\eeq
Our interest is in this quantity $\pi_n(L)$.
Eqs.\,(\ref{intropinL}) and (\ref{introQnL}) show how we may decompose it 
into the probability $p_n$ that the interface have $n$ edges 
and the conditional probability $Q_n(L)$ that the focal
distance associated with an $n$-edged face be equal to $L$.


\subsection{The probability $\pi_n(L)$ as a $3n$-fold integral}
\label{sec3nfold}

Since the probability $P$ is the same for all $(i,j)$,
we will take for definiteness $(i,j)=(N-1,N)$.
We can find an expression for $P_n(L)$ by writing $P$ as an integral over all
seed configurations and inserting the appropriate indicator function
$\chi_n(\bR_1,\ldots,\bR_{N-2};\bR_{N-1},\bR_N)$ 
which is unity if cells $N-1$ and $N$ share an $n$-edged
face and vanishes otherwise. We get
\beq
P = \frac{1}{V^N}\int_V\dd\bR_1\ldots\dd\bR_N\,\,\sum_{n=3}^\infty 
\chi_n(\bR_1,\ldots,\bR_{N-2};\bR_{N-1},\bR_N). 
\label{xpinitial}
\eeq
Obviously $\chi_n$ can depend on the seed positions
$\bR_{N-1}$ and $\bR_N$ only
through their distance $|\bR_{N-1}-\bR_N|\equiv 2L$.
We may therefore fix these seeds at 
$\bR_{N-1}=\bL_1\equiv(0,0,L)$ and $\bR_N=\bL_2\equiv(0,0,-L)$
while replacing the integration $\int_V \dd\bR_{N-1}\dd\bR_N$ by
$ 4\pi V \int_0^\infty(2L)^2\dd(2L)$.
Eq.\,(\ref{xpinitial}) then becomes
\beq
P = \frac{32\pi}{V^{N-1}}\int_V\dd\bR_1\ldots\dd\bR_{N-2}
\int\dd L\,\,L^2 \,\,\sum_{n=3}^\infty \chi_n(\bR_1,\ldots,\bR_{N-2};L),
\label{xpNm2}
\eeq
where the last argument in $\chi_n$ is now meant as a reminder that
$\bR_{N-1}=\bL_1$ and $\bR_N=\bL_2$.
The face shared by the two cells now lies in the $xy$ plane.
\vspace{3mm}

Each edge of the face is shared by the two focal cells and a third cell
that we will refer to as {\it first-neighbor cell}; we will call its seed a
{\it first-neighbor seed.}
There are $N-2 \choose n$ 
equivalent ways of choosing the $n$ first-neighbor seeds among
the $N-2$ seeds 
over whose positions we integrate in (\ref{xpNm2}).
By a permutation of indices we may take the first neighbors
to be those of
coordinates  $\bR_1,\bR_2,\ldots\bR_n$ and compensate by an extra factor 
$N-2 \choose n$ in the expression for $P$.

For each $j=1,2,\ldots,N-2$,
the two planes that perpendicularly bisect $L_1R_j$ and $L_2R_j$%
\footnote{We write $AB$ for the line segment connecting the points $\bA$
and $\bB$, and will write $\overline{AB}$ for its length.} 
also cut the $xy$ plane in a common line that we will call $\ell_j$.

We split $\chi_n$ according to
\beq
\chi_n(\bR_1,\ldots,\bR_{N-2};L) = \chi(\bR_1,\bR_2,\ldots\bR_n;L)
\prod_{j=n+1}^{N-2}\overline{\chi}(\bR_j|\bR_1,\ldots\bR_n;L)
\label{xchin}
\eeq
in which the first factor on the RHS contains the conditions on
$\bR_1,\ldots,\bR_n$, and the product the conditions on the
remaining seed positions; explicitly\\
${}$\phantom{i}(i)
$\chi(\bR_1,\ldots,\bR_n;L)$ is unity if the 
$\ell_m$ with $m=1,2,\ldots,n$ 
enclose a convex $n$-gon (which is then the face; see Fig.\,\ref{figure_1}), 
and is zero otherwise; 

(ii) $\overline{\chi}(\bR_j|\bR_1,\ldots\bR_n;L)$
is unity if the perpendicular bisecting plane of $L_1R_j$ 
(and hence also the one of $L_2R_j$)
intersects the $xy$ plane along a line $\ell_j$ that does not cut
the face. This will be true if and only if $\bR_j$
stays outside a three-dimensional domain $\pcV(\bR_1,\ldots,\bR_n;L)$  
determined uniquely by the positions of the two focal seeds and the $n$ 
first-neighbor seeds. We will find an explicit characterization of this
domain later; its volume will be denoted by $\cV$. 
\vspace{2mm}

\begin{figure}
\begin{center}
\scalebox{.50}
{\includegraphics{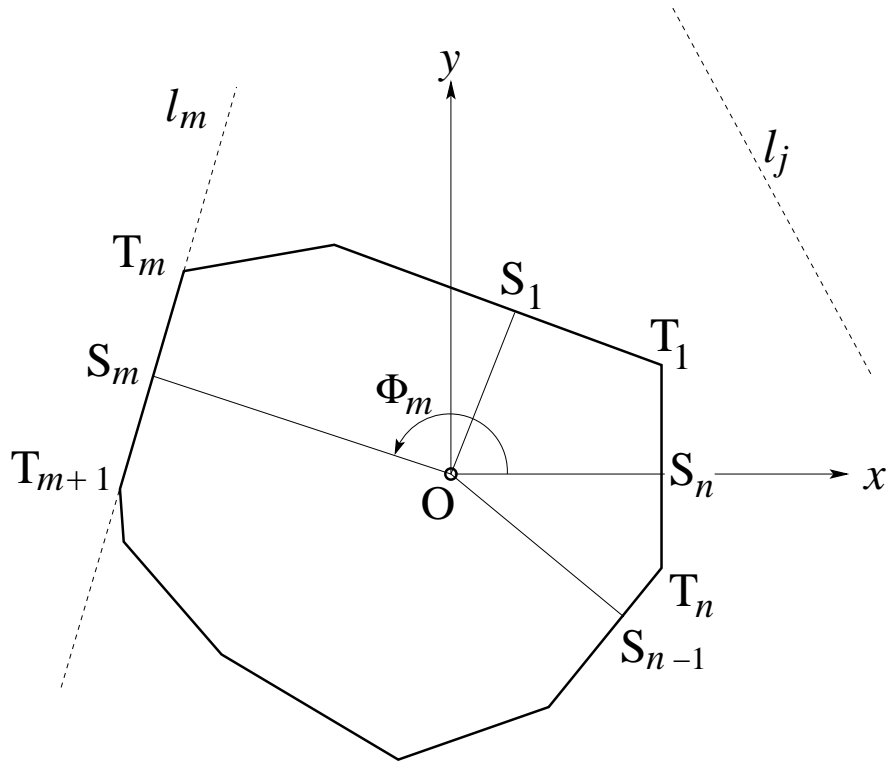}}
\end{center}
\caption{{\small Geometry in the $xy$ plane.
The $n$-sided face between two adjacent cells has vertices
$\bT_1,\bT_2,\ldots,\bT_n$. The lines $\ell_j$ for $j=1,2,\ldots,N-2$ 
are defined in section \ref{sec3nfold}.
The edges of the face lie on the $\ell_m$ with $m=1,2,\ldots,n$. 
The $\ell_j$ with $n+1,n+2,\ldots,N-2$ do not intersect the face. 
}}
\label{figure_1}
\end{figure}

When inserting (\ref{xchin}) in (\ref{xpNm2}) and integrating over
$\bR_{n+1},\ldots,\bR_{N-2}$ we get 
\bea
P &=& \frac{32\pi}{V^{N-1}} \int\dd L\,\,L^2 \,\,\sum_{n=3}^\infty
{N-2 \choose n}\int\dd\bR_1\ldots\dd\bR_n\,\,
\chi(\bR_1,\ldots,\bR_n;L)\,\,
\nonumber\\
&& \qquad \times\big[ V-\cV(\bR_1,\ldots,\bR_n;L) \big]^{N-2-n}.
\label{xpn}
\eea
Upon comparing (\ref{xpn}) and (\ref{dpnL}) we identify $P_n(L)$.
We now multiply both members of (\ref{xpn}) by $N$,
use (\ref{dpinL}) and (\ref{cp}), and take the limit
$N,V\to\infty$ at fixed $\lambda=N/V$. This leads to 
\beq
\pi_n(L) = \frac{32\pi\lambda^{n+1}}{c_F}\frac{L^2}{n!}
\int\dd\bR_1\ldots\dd\bR_n\,\
\chi(\bR_1,\ldots,\bR_n;L)\,\,
\ee^{-\lambda\cV(\bR_1,\ldots,\bR_n;L)}.
\label{xpinL}
\eeq
With equation (\ref{xpinL}) we have achieved expressing $\pi_n(L)$ as a
$3n$-fold integral. 

Before embarking upon the explicit evaluation of (\ref{xpinL})
we pass to spherical coordinates
$\bR_m=(R_m,\Phi_m,\Theta_m)$ defined as follows.
As usual, $R_m$ is the length of $\bR_m$ and $\Phi_m$ is the azimuthal angle,
measured with respect to the positive $x$ axis;
however, in deviation from standard usage, the polar angle $\Theta_m$ will be 
measured from the $xy$ plane
in the direction of the positive $z$ axis; that is, we have 
$-\frac{\pi}{2}\leq\Theta_m\leq\frac{\pi}{2}$.
This definition allows us 
to maintain an explicit symmetry between the two half-spaces above and below
the plane of the face.
Eq.\,(\ref{xpinL}) then transforms into
\bea
\pi_n(L) &=& \frac{32\pi\lambda^{n+1}}{c_F}\frac{L^2}{n!}
\int_0^\infty \prod_{m=1}^n \dd R_m\,R_m^2\,\,
\int_{-1}^1 \prod_{m=1}^n \dd\sin\Theta_m
\nonumber\\[2mm]
&& \times \int_0^{2\pi}  \prod_{m=1}^n \dd\Phi_m \,\,
\chi(\bR_1,\ldots,\bR_n;L)\,\,
\ee^{ -\lambda\cV(\bR_1,\ldots,\bR_n;L) }.
\label{xpinLxi}
\eea
We consider this $3n$-fold integral for $\pi_n(L)$ as the starting point 
of this paper. Our purpose will be to render it more
explicit and to extract from it the most interesting information that it contains.


\section{Coordinates in the $m$th first-neighbor plane}
\label{secintTheta}

\subsection{Geometry}
\label{secgeometry}

Let us suppose for convenience that 
$m=1,2,\ldots,n$ numbers the edges of the face in counterclockwise order.%
\footnote{This may be achieved by a permutation of the indices $1,2,\ldots,n$; see section \ref{secintegrationsPhim}.}
In the $xy$ plane, 
let $\bS_m$ be the projection of the origin $\bO$ onto $\ell_m$;
this point may lie on the $m$th edge of the face or on its extension
(see figure \ref{figure_1}). 
It is equidistant to the three seeds $\bL_1$, $\bL_2$, and $\bR_m$,
and lies in the plane passing through these seeds, 
which we will call the $m$th {\it first-neighbor plane\,}
(see figure \ref{figure_2}).  
We write $r_m$ for the radius of the circle of center $\bS_m$ that passes
through these three seeds.
Since the set of projections $\{\bS_m|m=1,2,\ldots,n\}$
completely determines the face, we will 
refer it as the set of {\it face coordinates}.

\begin{figure}
\begin{center}
\scalebox{.50}
{\includegraphics{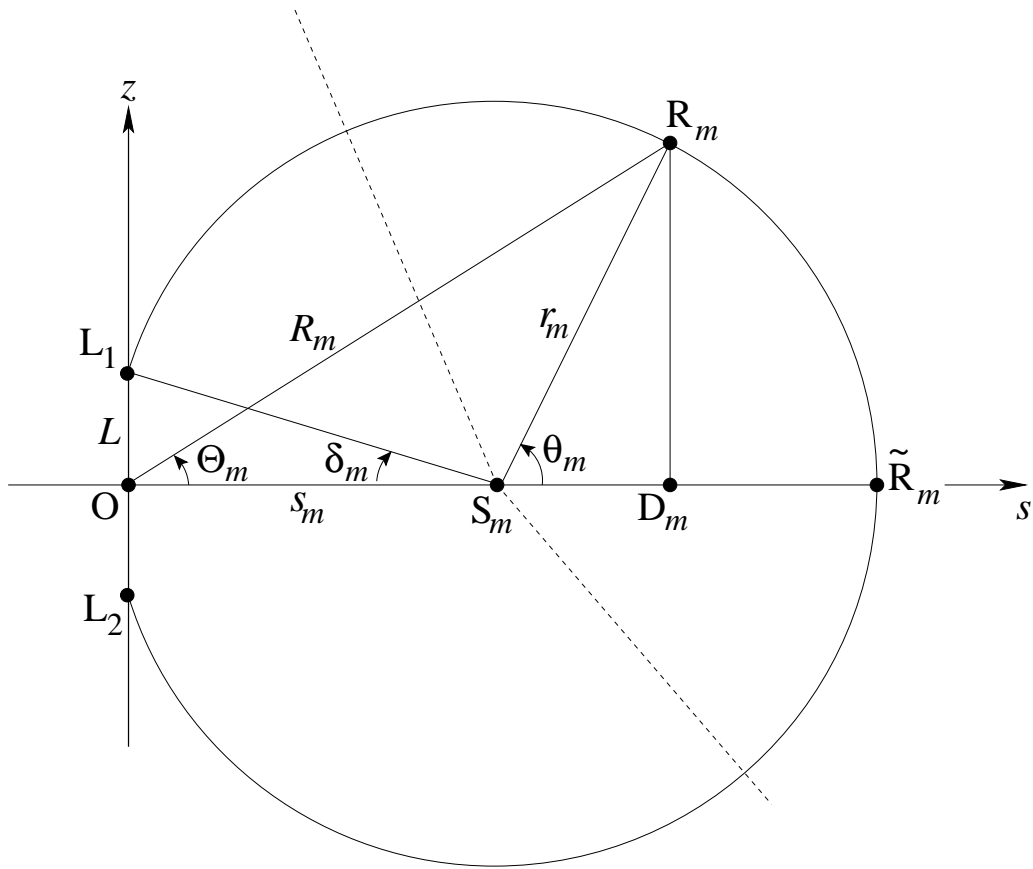}}
\end{center}
\caption{{\small The $m$th first-neighbor plane 
contains the seeds $\bL_1$, $\bL_2$, and $\bR_m$. 
The three corresponding cells share pairwise three faces whose intersections
with the plane of the figure lie on the two dashed lines 
(the perpendicular bisectors of $L_1R_m$ and $L_2R_m$)
and the $s$ axis (perpendicular bisector of $L_1L_2$). 
}}
\label{figure_2}
\end{figure}

\subsection{Coordinate transformation}
\label{secfirstneighbor}
 
In the $m$th first-neighbor plane the coordinates $R_m$ and $\Theta_m$
identify the $m$th first-neighbor seed. 
We will now prepare for integrating over these coordinates. To that end
we will transform them to coordinates $s_m$ and $\theta_m$ 
defined in figure \ref{figure_2}. We do this in two steps: first from 
$(R_m,\Theta_m)$  to  $(r_m,\theta_m)$ and then from $r_m$ to $s_m$.
 
\subsubsection{From $(R_m,\Theta_m)$  to  $(r_m,\theta_m)$}
\label{sectransf1}

To find the transformation we observe that $\overline{D_mR_m}$ 
may be calculated in the triangles $OD_mR_m$  and $C_mD_mR_m$, which yields
\beq
R_m\sin\Theta_m = r_m\sin\theta_m\,.
\label{trF1}
\eeq
Furthermore the relation $\overline{OD_m}=\overline{OC_m}+\overline{C_mD_m}$
may be expressed as
\bea
R_m\cos\Theta_m &=& \sqrt{r_m^2-L^2} + r_m\cos\theta_m \nonumber\\[2mm]
                &=& r_m[\cos\delta_m + \cos\theta_m],
\label{trF2}
\eea
where we introduced the abbreviation
\beq
\cos\delta_m = 
\sqrt{ 1-{L^2}/{r_m^2} }\,.
\label{defdeltam}
\eeq
From (\ref{trF1}) 
and (\ref{trF2}) one deduces for $R_m^2$ and $\sin^2\Theta_m$ the following
explicit expressions in terms of the new coordinates
\bea
R_m^2 &=& 2r_m^2 - L^2 +2r_m^2\cos\delta_m\cos\theta_m\,,
\nonumber\\[2mm]
\sin^2\Theta_m &=& \frac{r_m^2\sin^2\theta_m}
{2r_m^2 - L^2 +2r_m^2\cos\delta_m\cos\theta_m}\,.
\label{xRThrth}
\eea
We will need the Jacobian  
$J_m = \p(R_m,\sin\Theta_m)/\p(r_m,\sin\theta_m)$
of this transformation.


\subsubsection{The Jacobian $J_m$}
\label{secJacobian}

To calculate the Jacobian $J_m$ we abbreviate 
\beq
R_{mr}=\frac{\p R_m}{\p r_m}\,,\ \
R_{m\theta}=\frac{\p R_m}{\p\sin\theta_m}\,,\ \
\Theta_{mr}=\frac{\p\sin\Theta_m}{\p r_m}\,,\ \
\Theta_{m\theta}=\frac{\p\sin\Theta_m}{\p\sin\theta_m}\,.  
\eeq
Upon deriving Eqs.\,(\ref{trF1}) and (\ref{trF2}) with respect to $r_m$ and 
$\sin\theta_m$ we get the four equations
\bea
R_{mr} \sin\Theta_m + \Theta_{mr} R_m &=& \sin\theta_m\,,\nonumber\\[2mm]
R_{mr} \cos\Theta_m - \Theta_{mr} R_m\tan\Theta_m  &=& 
\frac{1}{\sqrt{1-L^2/r_m^2}} + \cos\theta_m\,,
\label{firstpair} \\[2mm]
R_{m\theta} \sin\Theta_m + \Theta_{m\theta} R_m &=& r_m\,,\nonumber\\[2mm]
R_{m\theta} \cos\Theta_m - \Theta_{m\theta} R_m\tan\Theta_m &=& 
-r_m\tan\theta_m\,.
\label{secondpair}
\eea
We may solve $R_{mr}$ and $\Theta_{mr}$ from Eqs.\,(\ref{firstpair})
and $R_{m\theta}$ and $\Theta_{m\theta}$ from Eqs.\,(\ref{secondpair}).
After some algebra this yields
\bea
J_m &=& R_{mr}\Theta_{m\theta}-R_{m\theta}\Theta_{mr} \nonumber\\[2mm]
    &=& \frac{r_m\cos\Theta_m}{R_m}
\left[ \frac{1}{\cos\theta_m}+\frac{1}{\sqrt{1-L^2/r_m^2}} \right]
\nonumber\\[2mm]
    &=&  \frac{\cos^2\Theta_m}{\cos\theta_m\cos\delta_m}\,,
\label{calcJ}
\eea
where to arrive at the last line we used (\ref{trF2}).

In the integral in (\ref{xpinLxi}) this change of variables of integration
therefore leads to the transformation
\bea
\dd R_m\,R_m^2\,\dd\sin\Theta_m &=& \dd r_m\,\dd\sin\theta_m\,J_mR_m^2
\nonumber\\[2mm]
&=& \dd r_m\,r_m^2\,\,\dd\sin\theta_m\,\,
\frac{(\cos\delta_m+\cos\theta_m)^2}{\cos\delta_m\cos\theta_m}
\nonumber\\[2mm]
&=& \dd r_m\,r_m^2\,\,\dd\theta_m\,\,
\left( 1-\frac{L^2}{r_m^2} \right)^{-\frac{1}{2}} 
\left[ \left(  1-\frac{L^2}{r_m^2} \right)^{\frac{1}{2}} 
+ \cos\theta_m \right]^2,
\nonumber\\
&&
\label{elint}
\eea
where to pass from the first to the second line we used (\ref{trF2}) and 
(\ref{calcJ}).


\subsubsection{From $r_m$ to $s_m$}
\label{sectransf2}

We set
\beq
s_m^2 = {r_m^2-L^2}
\label{dsm}
\eeq
and refer again to figure \ref{figure_2}.
We then find from Eq.\,(\ref{elint}) that
\beq
\dd R_m\,R_m^2\,\dd\sin\Theta_m = \dd s_m\, s_m^2\,\dd\theta_m\,\,
\left[ 1\,+\, 
\frac{\cos\theta_m}{\cos\delta_m} \right]^2
\label{elints}
\eeq
where now
\beq
\frac{1}{\cos\delta_m} = \sqrt{1+\frac{L^2}{s_m^2}}\,.
\label{rcosdeltas}
\eeq
It may be noted that for $\delta_m\to\pi/2$ the divergence of the bracketed
factor in (\ref{elints}) is compensated by the fact that in that limit 
$s_m\to 0$. 


\subsubsection{Range of $s_m$ and $\theta_m$}
\label{secrange}

When the first-neighbor position $\bR_m$ 
is integrated over the $m$th first-neighbor half-plane
(that is, at fixed angle $\Phi_m$),
it will also run through 
the half-disk of center $\bO$ and radius $L$ that is part of this half-plane.
In that case the center $\bS_m$ moves into the complementary half-plane, which
we may express by letting the coordinate $s_m$ (see figure \ref{figure_2})
be the negative square root of (\ref{dsm})
and $\delta_m$ the angle between the negative $s$ axis and $OL_1$.
This corresponds to the origin $\bO$ lying outside the cell face
and to the azimuthal angle of $\bS_m$ being equal to $\Phi_m+\pi$.
For large $n$ the relative weight 
of this special subclass of faces
will be exponentially small in $n$, and therefore
negligible once we expand in powers of $n$ in sections 
\ref{seclargenexpansion} through \ref{secarriving}. 
It will be convenient to suppress this subclass from here on and let 
the new variables $s_m$ and $\theta_m$ range through
\beq
0<s_m<\infty, \qquad -\pi+\delta_m < \theta_m < \pi-\delta_m\,,
\label{limrth}
\eeq
with $\delta_m$ given in terms of $s_m$ by (\ref{rcosdeltas}).
Eq.\,(\ref{xpinLxi}) may then be rewritten as
\bea
\pi_n(L) &\simeq& \frac{32\pi\lambda^{n+1}}{c_F}\frac{L^2}{n!}
\int_{0}^\infty \prod_{m=1}^n \dd s_m\,s_m^2 
\int \prod_{m=1}^n \dd\theta_m\,\,
\Big[ \,1\,+\,\frac{\cos\theta_m}{\cos\delta_m} \Big]^2 
\nonumber\\[2mm]
&& \times\int_0^{2\pi} \prod_{m=1}^n \dd\Phi_m\,\,
\chi(\bR_1,\ldots,\bR_n;L)\,\,
\ee^{-\lambda\cV(\bR_1,\ldots,\bR_n;L)},
\label{xpinLrth}
\eea
in which the limits of the $\theta_m$ integrations, not explicitly indicated,
are those of Eq.\,(\ref{limrth}); and where $\delta_m$ is given by
(\ref{rcosdeltas}). 

We will now show that $\chi$ and $\cV$ are independent of the angles
$\theta_m$ and that therefore the $\theta_m$ integrations in (\ref{xpinLrth})
are mutually independent and may be carried out fully explicitly.\\

\begin{figure}
\begin{center}
\scalebox{.40}
{\includegraphics{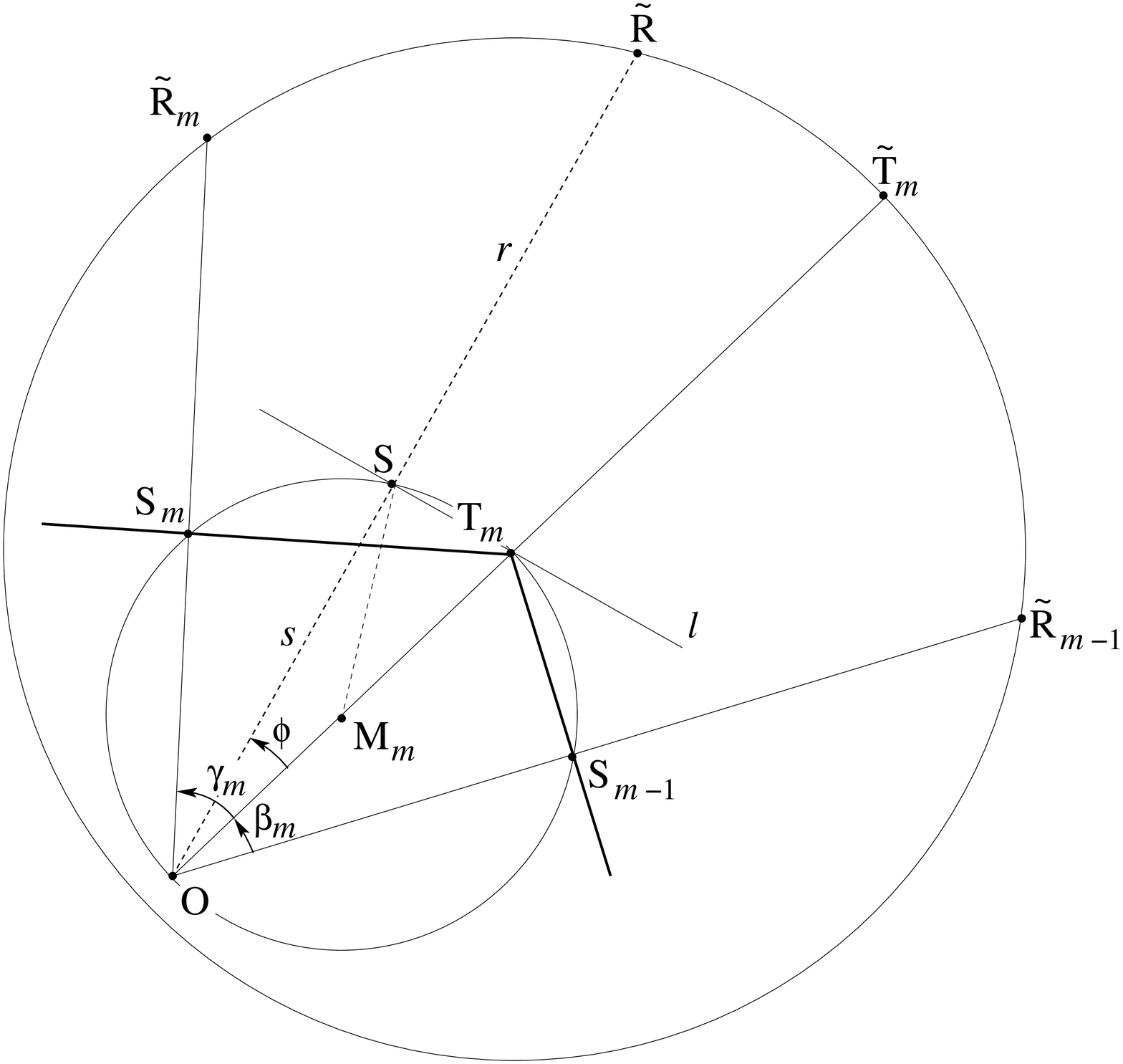}}
\end{center}
\caption{{\small 
More geometry in the $xy$ plane.
The large circle has $\bT_m$ and the small one $\bM_m$ as its center.
This figure defines the angles $\beta_m$, $\gamma_m$, and $\phi$. 
We define $\xi_m=\beta_m+\gamma_m$ and
$\eta_m=\gamma_m+\beta_{m+1}$ ($n$-periodicity understood).
When $\bS_{m-1}$ (or $\bS_m$) lies on the extension of the $(m-1)$th 
(or the $m$th) edge beyond the point $\bT_m$, then $\beta_m$ (or
$\gamma_m$) is negative. 
When $\phi$ varies from $-\beta_m$ to $\gamma_m$,
point $\bS$ moves along the smaller circle 
from $\bS_{m-1}$ to $\bS_m$ and
line $\ell$ pivots around $\bT_m$;
$\btR$ moves along the large circle from $\btR_{m-1}$ to $\btR_m$.
}}
\label{figure_3}
\end{figure}


\section{The excluded domain ${\pmb{\cV}}$}
\label{secexcludedvolume}

For given focal seeds in $\bL_{1,2}=(0,0,\pm L)$ and a given set of 
first-neighbor positions $\{\bR_1,\ldots,$ $\bR_n\}$,
the domain $\pcV(\bR_1,\ldots,$$\bR_n;L)$ 
is the region of space from which
the remaining seeds are excluded if they are not to interfere with
the first-neighbor relations. If one of those remaining 
seeds entered the excluded domain, it would itself become a
first neighbor, contrary to what had been supposed.
In this section we will obtain
an explicit characterization of the excluded domain
$\pcV(\bR_1,\ldots,\bR_n;L)$ .  
It will apear that it in fact depends only on
the more restricted set of face coordinates $\{\bS_1,\ldots,\bS_n\}$,
that are all located in the $xy$ plane. 
This feature will allow us to explicitly carry out the integrations over the
angles $\theta_m$. 




\subsection{Geometry in the half-plane at angle $\Phi$}

Let $\bR^\prime$ be the position of an arbitrary one of
the remaining seeds $\bR_j$ (where $j=n+1,\ldots,N-2$)
and let $\Phi$ be its azimuthal angle.
Let furthermore$\ell_j$ be the line along which
the perpendicular bisecting planes of $L_1R^\prime$ and of $L_2R^\prime$
intersect the $xy$ plane; and let $\bS^\prime$
be the projection of the origin $\bO$ onto $\ell^\prime$.

The considerations that follow all concern the vertical
half-plane passing through
the $z$ axis and $\bR^\prime$,
and that we will refer to as the {\it half-plane at angle $\Phi$}.
This half-plane contains $\bS^\prime$.

We have the following property.
Let $r^\prime\equiv \overline{S^{\prime}R^\prime}$.
{\it When in this half-plane $\bR^\prime$ moves 
along a circular arc of center $\bS^\prime$ and radius $r^\prime$,
having its end points in $\bL_1$ and $\bL_2$\,,
then $\ell^\prime$ remains invariant.}
That is, in this half-plane 
this circular arc is a locus of positions $\bR^\prime$
that are equivalent in the sense of leading to the same $\ell^\prime$.
We will refer to this arc as arc $L_1R^\prime L_2$ and   
denote its intersection with the $xy$ plane by $\btR^\prime$. 
\vspace{2mm}

We wish to investigate under which conditions
$\btR^\prime$ (and therefore any other point on the circular arc
to which it belongs) is such that $\ell^\prime$
does not cut the $n$-edged face; 
under these conditions $\btR^\prime$  is outside the excluded domain $\pcV$. 

Suppose $\ell^\prime$ cuts the face.
Let us then move seed $\bR^\prime$ in such a way that
the bisector plane of 
$L_1\tR$ moves parallel to itself away from $\bL_1$.%
\footnote{Moving, instead, the bisector plane of $L_2\tR$ away from $\bL_2$
  would lead to the same conclusions.} 
This means that in the $xy$ plane along the half-axis at angle $\Phi$ the 
position $\btR^\prime$ and the center $\bS^\prime$
move away from the origin while being related by
\beq
L^2 + {s^\prime}^2={r^\prime}^2, \qquad \tR^\prime = s^\prime + r^\prime,
\label{circarcprime}
\eeq
in which $s^\prime =\overline{OS^\prime}$ and 
$r^\prime = \overline{S^\prime\tR^\prime}=\overline{S^\prime R^\prime}$. 
Line $\ell^\prime$
will stop cutting the face when it passes only through
a single vertex; 
if $\Phi$ is in the sector $\Phi_{m-1}<\Phi<\Phi_m$\,, 
this will be vertex $\bT_{m}$. This is the
situation is represented in figure \ref{figure_3}, where 
we have denoted by  $\ell, \bS$, and $\btR$ (without primes)
the positions then occupied by
$\ell^\prime, \bS^\prime$, and $\btR^\prime$, respectively.
Setting $s\equiv \overline{OS}$ and $r\equiv \overline{S\tR}$
we have from Eq.\,(\ref{circarcprime}) that
$s$ and $r$ are related by
\beq
L^2 + s^2 = r^2, \qquad \tR = s+r.
\label{circarc}
\eeq
The two equations (\ref{circarc}) contain three unknowns
$s$, $r$, and $\tR$ that we would like to
determine in terms of the running angle $\Phi$
and the face coordinates $\{\bS_1,\ldots,\bS_n\}$.
The third equation comes from the condition that $\ell$ pass through 
$\bT_m$. We use the angles $\beta_m\,, \gamma_m\,,$ and $\phi$
defined in figure \ref{figure_3}.
Here the auxiliary angle $\phi\equiv\Phi-\Phi_m$ 
is the ``local'' azimuthal angle in the $m$th sector and has the range 
\beq
 -\beta_m<\phi<\gamma_m\,.
\label{rangephi}
\eeq
Since $s=\overline{OT_m}\cos\phi$ 
and since for $\overline{OT_m}$ we may use either of the two expressions
$\overline{OT_m} = {s_{m-1}}/{\cos\beta_m} = {s_m}/{\cos\gamma_m}$\,,
it follows that
\beq
s(\phi) = {s_{m-1}}\frac{\cos\phi}{\cos\beta_m} 
  = {s_m}    \frac{\cos\phi}{\cos\gamma_m}\,.
\label{relsphi}
\eeq
The first one of Eqs.\,(\ref{circarc}) then gives $r(\phi)$ as
\beq
r(\phi) = s_{m-1} \left( \frac{\cos^2\phi}{\cos^2\beta_m}
+ \frac{L^2}{s_{m-1}^2} \right)^{1/2}
=
s_{m} \left( \frac{\cos^2\phi}{\cos^2\gamma_m}
+ \frac{L^2}{s_{m}^2} \right)^{1/2}.
\label{relrphi}
\eeq
Eqs.\,(\ref{relsphi}) and (\ref{relrphi}) are both valid in the $m$th sector,
that is, for $\Phi_{m-1}<\Phi<\Phi_m$, or equivalently, for $\phi$ in the 
range (\ref{rangephi}).

The functions $r(\phi)$ and $s(\phi)$ have hereby been
expressed entirely in terms of the 
coordinates that determine the face.
They have been defined sectorwise and
at the sector boundaries they are continuous with discontinuous derivatives. 
After these preliminaries it is now easy to show how they determine the excluded
domain. 


\subsection{Excluded domain $\pmb{\cV}$: a pumpkin}
\label{secvolumepumpkin}

In the vertical half-plane at angle $\Phi$, 
arc $L_1\tR L_2$ (of center $\bS$)
together with chord ${L_1L_2}$ encloses a truncated disk. 
When $\Phi$ is varied, this truncated disk 
sweeps out the excluded domain $\pcV$, 
that we will now be able to characterize.

When $\Phi$ varies within the $m$th sector 
(and hence the local angle $\phi$ varies from $-\beta_m$ to $\gamma_m$),
arc $L_1\tR L_2$ slides along 
a sphere that has its center in
$\bT_m$ and whose squared radius is $\overline{OT_m}^2+L^2$.
To see this, it suffices to note that 
$\bS$ is the center of the arc; $T_mS$ is perpendicular
to the plane of the arc and hence equidistant to all of its points; and
since $\bL_1$ is one of these points, that distance is   
$(\overline{OT_m}^2+L^2)^{1/2}$ and independent of $S$; 
hence all arcs have the same distance to $\bT_m$.
The surface $\partial\pmb{\cV}$
of the excluded domain is therefore piecewise spherical
and {\it the domain $\pmb{\cV}$ itself is the union of $n$ balls 
having their centers in the vertices $\bT_1,\ldots,\bT_n$ of the face.} 
\vspace{2mm}

In two dimensions, the excluded domain associated with a 2D Voronoi cell
is the union of $n$ disks and is often called the 
{\it Voronoi flower\,} of that cell.
In the present case the cell face has associated with it a 3D excluded
domain $\pcV$ which is the union of $n$ balls having their centers 
in the $xy$ plane; and because of the shape of its surface $\partial\pcV$ this
domain rightfully deserves the name {\it Voronoi pumpkin.}
Its intersection with the $xy$ plane is similar to a flower:
it is the union of $n$ disks centered on the vertices of the face.


\subsection{Large-$n$ limit of $\pmb{\cV}$:  a spindle torus}

This is too early a stage to take the large-$n$ limit.
However, it is now possible for us to
look ahead and guess what to expect.
 
In view of our experience with the 2D Poisson-Voronoi cell it is
reasonable to assume that for $n\to\infty$ the vertices $\bT_m$
become dense on a curve that tends towards a circle.
In that case the union of balls that constitute the excluded volume will
tend to a torus; the major radius of this torus cannot be larger than its
minor radius, but we cannot be sure at this point how either will 
scale with $n$. Plausible heuristic arguments presented elsewhere 
\cite{HilhorstLazar14} indicate that the volume of this torus
will approach $n/\lambda$.%
\footnote{A torus whose major radius is smaller than
  its minor radius is referred to as a {\it spindle torus}. 
Its surface is sometimes called an {\it apple.}}
It will appear at the end of our calculation that in fact
for $n\to\infty$ both radii are $\sim n^{1/3}$ and asymptotically equal 
to leading order. 


\subsection{Volume $\cV$ of the excluded domain $\pmb{\cV}$}
\label{secvolume}

For $n\to\infty$ the variation of $\Phi$ within a sector is of the
order of $\sim 2\pi/n$ and we may neglect the 
$\phi$ dependence of $r(\phi)$ and $s(\phi)$ 
within that sector. 
The volume enclosed between the two vertical half-planes
(at angles say $\Phi_{m-1}$ and $\Phi_m$) defining the sector, and the sphere
centered at $\bT_m$, is then an infinitesimal slice of a spindle torus
with major and minor radius equal to $s(\phi)$ and $r(\phi)$, respectively,
where $s(\phi)<r(\phi)$.
The volume of the excluded domain is therefore the sum of the volumes of these
slices. 

The volume $\Vsp(R_{_{+}},R_{_{-}})$ of a spindle torus with major radius
$R_{_{+}}$ and minor radius $R_{_{-}}$ is given by
\beq
\Vsp(R_{_{+}},R_{_{-}}) = 2\pi^2 R_{_{-}}^3 g(x), 
\qquad x^2\equiv 1-R_{_{+}}^2/R_-^2\,,
\label{Vspindletorus}
\eeq 
in which
\beq
g(x) = \frac{1}{\pi}\left[ x - \tfrac{1}{3}x^3
+ (\pi-\arcsin x)\sqrt{1-x^2} \right].
\label{xgx}
\eeq
In the present case we have $R_{_{+}}=r(\phi)$ and $R_{_{-}}=s(\phi)$ whence
\beq
x^2(\phi) \equiv 1-\frac{s^2(\phi)}{r^2(\phi)}
= \frac{L^2}{r^2(\phi)}\,.
\label{defxm}
\eeq
The infinitesimal toroidal slice 
swept out therefore has a volume
$2\pi^2 r^3(\phi)g(x(\phi))\times(\Delta\Phi/2\pi)$,
and this should be integrated over $\Phi$ to
yield the excluded volume $\cV$.
This gives
\beq
\cV=\frac{1}{2\pi}\sum_{m=1}^n \int_{-\beta_m}^{\gamma_m}\dd\phi\,\,
2\pi^2 r^2(\phi)g(x(\phi)),
\label{VsumV}
\eeq
it being understood here and henceforth
that under the sum on $m$ the functions $r(\phi)$ and $x(\phi)$ take their
expressions valid in the $m$th angular sector.
Eq.\,(\ref{VsumV}) together with the substitutions 
(\ref{relrphi}), (\ref{xgx}), and (\ref{defxm})
yields the excluded volume in terms of the face coordinates.
The $m$th term in the sum in (\ref{VsumV}) depends on the
face coordinates $s_{m-1},s_m,\beta_m,$ and $\gamma_m$. 

We have shown, therefore, that $\pcV$ depends only on the coordinates 
$\{\bS_1,\ldots,\bS_n\}$ and on $L$. The same remark holds for
the indicator $\chi$.
We will therefore write these two functions from here on as
$\pcV(\bS_1,\ldots,\bS_n;L)$ and $\chi(\bS_1,\ldots,\bS_n;L)$.
That is, they are independent of the polar angles $\theta_m$\,.
We are now able to do the integrations on the angles $\theta_m$.


\section{Integrating over the polar angles $\theta_m$}
\label{secintegratingtheta}

We take up again the calculation of $\pi_n(L)$, for which we found 
expression (\ref{xpinLrth}). We now exploit the fact just shown 
that $\cV$ and $\chi$ are independent of the polar angles $\theta_m$.
The polar angle integrations in (\ref{xpinLrth}) 
therefore factorize and the one on $\theta_m$ becomes
\bea
\int_{-\pi+\delta_m}^{\pi-\delta_m}\dd\theta_m\,
\left[\,1\,+\,\frac{\cos\theta_m}{\cos\delta_m}\right]^2 
&=& \frac{(\pi-\delta_m)(3-2\sin^2\delta_m)}{\cos^2\delta_m} 
\,+\,3\tan\delta_m 
\nonumber\\[2mm]
&=& 3\pi + \pi\tan^2\delta_m + 3\big( \tan\delta_m-\delta_m \big)
-\delta_m\tan^2\delta_m \nonumber\\[2mm]
&\equiv& 3\pi K\left( \frac{L}{s_m} \right),
\label{xKm}
\eea
in which $\cos\delta_m$ is given in terms of $L/s_m$ by (\ref{rcosdeltas})
and where the last line defines $K(x)$, which is such that $K(0)=1$.
We will set
\beq
\ee^{\cal K} = \prod_{m=1}^n K\left( \frac{L}{s_m} \right) .
\label{defcalK}
\eeq
Upon using (\ref{xKm}) and (\ref{defcalK})
in (\ref{xpinLrth}) we find for $\pi_n(L)$ the expression
\bea
\pi_n(L) &\simeq& \frac{32\pi\lambda^{n+1}}{c_F}\frac{(3\pi)^nL^2}{n!}
\int_{0}^\infty \left[ \prod_{m=1}^n \dd s_m\,s_m^2 \right]
\ee^{\cal K}
\nonumber\\[2mm]
&& \times \int_0^{2\pi}\prod_{m=1}^n\dd\Phi_m
\,\,\chi(\bS_1,\ldots,\bS_n;L)\,\,
\ee^{-\lambda\cV(\bS_1,\ldots,\bS_n;L)}.
\label{xpinLk}
\eea
The problem of calculating $\pi_n(L)$ has hereby been reduced from $3n$ to
$2n$ coupled integrations: $n$ radial and $n$ angular ones in the $xy$ plane. 
In the next sections we will subject
Eq.\,(\ref{xpinLk}) to several further transformations,
the purpose being to cast it in a form amenable to a large-$n$ expansion.


\section{Rewriting the $2n$ integrations in the $xy$ plane}
\label{secrewriting2n}


\subsection{Transforming the radial integrations}
\label{secradialintegrations}


We are motivated by the idea that for any
given $n$-edged face, in the large-$n$ limit the $s_m$
will tend to be close to their $m$-averaged value.
We therefore define the {\it face radius\,} $\sav$ and 
relative ``radii'' $\sigma_m$ by
\beq
\sav = \frac{1}{n}\sum_{m=1}^n s_m  \quad\mbox{ and } \quad 
s_m=\sav\sigma_m\,, \quad m=1,2,\ldots,n.
\label{dravrhom}
\eeq
Note that the $m$-average $\sav$ 
is still a random quantity that varies from one face to another.
For later use we also define variables $\tau_m$ by
\beq
\sigma_m=1+\tau_m\,.
\label{dtaum}
\eeq
The $\sigma_m$ and $\tau_m$ satisfy 
the two equivalent sum rules
\beq
 \sum_{m=1}^n\sigma_m=1, \qquad  \sum_{m=1}^n\tau_m=0.
\label{sumrulesigmatau}
\eeq
When $n$ gets large we expect the $\sigma_m$ to be close to unity and
therefore the $\tau_m$ small in a way to be specified quantitatively later.

We may now pass in (\ref{xpinLk})
from the $n$ integrations on the $s_m$ to a single integration on
$\sav$ and $n$ integrations on the $\sigma_m$, while introducing a delta
function constraint that enforces sum rule (\ref{sumrulesigmatau}).
This converts Eq.\,(\ref{xpinLk}) into
\bea
\pi_n(L) &\simeq& \frac{32\pi\lambda^{n+1}}{c_F}\frac{(3\pi)^nL^2}{n!}
\int_{0}^\infty\dd\sav\,\sav^{3n-1}
\nonumber\\[2mm]
&&\times \int_{0}^\infty \left[
\prod_{m=1}^n \dd\sigma_m\,\sigma_m^2\,
K\left( \frac{L}{\sav\sigma_m} \right) \right]
\delta\left( 1-\frac{1}{n}\sum_{m=1}^n\sigma_m \right)
\nonumber\\[2mm]
&& \times 
\int_0^{2\pi} \prod_{m=1}^n \dd\Phi_m\,
\chi(\bS_1,\ldots,\bS_n;L)\,\,
\ee^{-\lambda\cV(\bS_1,\ldots,\bS_n;L)}.
\label{xpinLrhom}
\eea
We will now start working on the angular integrations.


\subsection{Transforming the angular integrations}
\label{secintegrationsPhim}

We transform the $\Phi_m$ integrations in (\ref{xpinLrhom})
in a succession of three steps, largely identical to the procedure followed in 
Ref.\,\cite{Hilhorst05b}. We therefore indicate these steps only succinctly.
 
First, we may choose in (\ref{xpinLrhom}) one of the angles $\Phi_m$ to be zero, say $\Phi_n=0$,
if we multiply the RHS by $2\pi$ to compensate, and we may order the angles
according to $0<\Phi_1<\Phi_2<\ldots<\Phi_{n-1}<2\pi$
if we multiply the RHS by $(n-1)!$ to compensate. This ordering 
makes it convenient to pass from the $\Phi_m$
to the angular differences $\xi_m$ (with $m=1,\ldots.n$) defined by
\bea
&&\xi_m = \Phi_m - \Phi_{m-1}\,, \qquad m=1,\ldots,n-1,\nonumber\\
&&\xi_n = 2\pi - \Phi_{n-1}\,.
\label{defxim}
\eea
The geometry imposes the constraints $0<\xi_m<\pi$ as well as the sum rule 
$\sum_{m=1}^{n}\xi_m=2\pi$.
\vspace{1mm}

In the second step we transform for each $m=1,2,\ldots,n$ separately
from the angle $\xi_m$ to the angle
$\beta_m$ defined in figure \ref{figure_3}.
We observe that
\beq
\sigma_m = \frac{\cos\gamma_m}{\cos\beta_m}\,\sigma_{m-1}\,,
\qquad m=1,2,\ldots,n, \quad \sigma_0\equiv\sigma_n\,,
\label{recrhom}
\eeq
which allows us, 
under the $\sigma_m$ integrations in Eq.\,(\ref{xpinLrhom}), 
to view $\gamma_m$ as a function of $\beta_m$.
Hence we have $\xi_m\equiv\beta_m+\gamma_m
=\beta_m+\arccos[(\sigma_m/\sigma_{m-1})\cos\beta_m]$.
The transformation from $\xi_m$ to $\beta_m$ is therefore
accompanied by a Jacobian 
$j_m\equiv\dd\xi_m/\dd\beta_m  = \sin(\beta_m+\gamma_m)/[\cos\beta_m\sin\gamma_m]$.
The $\beta_m$ integrals must be appropriately nested.
The geometrical condition that the face vertices $\bT_m$
have azimuthal angles that increase with $m$ is expressed by
the condition $\beta_{m+1}>-\gamma_m$. 
This is equivalent to imposing the condition expressed by
the indicator function $\chi$, which may therefore from here on
be omitted. Periodicity is ensured by a factor
$\theta(\gamma_n+\beta_1)$, where $\theta$ is the Heaviside function.
\vspace{1mm}

Thirdly, we introduce extra integrations over the $n$ variables $\gamma_n$,
compensated by the introduction of a product of Dirac delta functions
that impose the $n$ relations (\ref{recrhom}).
\vspace{1mm}

Carrying these three steps out leads to
\bea
\int_0^{2\pi}\prod_{m=1}^n\dd\Phi_m &=&
2\pi(n-1)!\int_0^\pi\prod_{m=1}^n\dd\xi_m\,
\delta\big( \sum_{m=1}^{n}\xi_m-2\pi \big) 
\chi(\bS_1,\ldots,\bS_n;L)
\nonumber\\
&& \\
&=&
2\pi(n-1)!\int_{-\pi/2}^{\pi/2}\!\!\dd\beta_1
\int_{-\gamma_{_1}}^{\pi/2}\!\!\dd\beta_2\ldots
\int_{-\gamma_{_{n-1}}}^{\pi/2}\!\!\dd\beta_n\, \theta(\gamma_n+\beta_1) 
\nonumber\\[2mm]
&& \times\,\,
\,j_1 j_2 \ldots j_n\,
\delta\big(\sum_{m=1}^n(\beta_m+\gamma_m)-2\pi\big) \\[2mm]
&=&2\pi(n-1)!\int\dd\beta\dd\gamma\,\,
\left[ \prod_{m=1}^n t_m \sigma_{m-1}\,
\delta\big(\sigma_m-\frac{\cos\gamma_m}{\cos\beta_m}\sigma_{m-1}\big) 
\right],
\nonumber\\
&&
\label{trfPhiint}
\eea
where we define
\beq
t_m=\frac{\sin(\beta_m+\gamma_m)}{\cos\beta_m\cos\gamma_m}
\label{defTm}
\eeq
and employ the shorthand notation
\bea
\int\dd\beta\dd\gamma &=&
\int_{-\pi/2}^{\pi/2}\!\!\dd\beta_1
\int_{-\beta_{_1}}^{\pi/2}\!\!\dd\gamma_1 
\int_{-\gamma_{_1}}^{\pi/2}\!\!\dd\beta_2
\int_{-\beta_{_2}}^{\pi/2}\!\!\dd\gamma_2 
\ldots\int_{-\gamma_{_{n-1}}}^{\pi/2}\!\!\dd\beta_n 
\int_{-\beta_{_n}}^{\pi/2}\!\!\dd\gamma_n \nonumber\\
&& \times\,\,     
\theta(\gamma_n+\beta_1)\,
\delta\big(\sum_{m=1}^n(\beta_m+\gamma_m)-2\pi\big). 
\label{defintbetagamma}
\eea
Eq.\,(\ref{trfPhiint}) is to be used to eliminate the $\Phi_m$ integrations in
(\ref{xpinLrhom}) in favor of the $\beta,\gamma$ integrations.
Whereas this certainly does not look like a simplification, it is a necessary
passage point on the way to our goal.


\subsection{Integrating over the radial variables $\sigma_m$}
\label{secintegratingsigmam}

Eq.\,(\ref{recrhom}) relates $\sigma_m$ to $\sigma_{m-1}$.
When iterating $m$ times we obtain
\beq
\sigma_m = \frac{\cos\gamma_m\cos\gamma_{m-1}\ldots\cos\gamma_1}
{\cos\beta_m\cos\beta_{m-1}\ldots\cos\beta_1}\,
\sigma_n, \qquad m=1,2,\ldots,n-1,
\label{recrhomit}
\eeq
where we recall the convention that $\sigma_0=\sigma_n$.
Let the function $G$ be defined by
\beq
G(\beta,\gamma)=\frac{1}{2\pi}\sum_{m=1}^n\big(\log\cos\gamma_m
                                              -\log\cos\beta_m\big).
\label{dG}
\eeq
Equation (\ref{recrhomit}) is valid also when we set $m=n$ and then
amounts to a condition on the angles that we have called the
the {\it no-spiral constraint}\,%
\footnote{The geometrical interpretation leading to this name was given
in Ref.\,\cite{Hilhorst05b}.}
and that may be expressed as
\beq
G(\beta,\gamma)=0
\label{nospiral}
\eeq
with $(\beta,\gamma)$ standing for the set of angles $\{\beta_m,\gamma_m|m=1,\ldots,n\}$.
In equation (\ref{xpinLrhom}) we now replace the integration on the angles
$\Phi_m$ by expression (\ref{trfPhiint}), interchange the integrations on the
$\beta_m$ and $\gamma_m$ with those on the $\sigma_m$, and carry out the
integrals on $\sigma_1,\sigma_2,\ldots,\sigma_{n-1}$ with the aid of the delta
functions. 
The integration on $\sigma_n$, finally, is easily carried out due to the delta
function constraint $\delta\big( 1-\frac{1}{n}\sum_{m=1}^n\sigma_m \big)$ 
in (\ref{xpinLrhom}).
The result of all this is that equation (\ref{xpinLrhom}) becomes
\bea
\pi_n(L) &\simeq& \frac{32\pi\lambda^{n+1}}{c_F}\frac{(3\pi)^nL^2}{n!}
\int_{0}^\infty\dd\sav\,\sav^{3n-1}
\nonumber\\[2mm]
&& \times\, 2\pi(n-1)!\int\dd\beta\dd\gamma\,\, \frac{\delta(G)}{2\pi}
\left[ \prod_{m=1}^n \sigma_m^3 t_m \right]
\,\ee^{{\cal K}(\sav,\beta,\gamma;L)-\lambda\cV(\sav,\beta,\gamma;L)} 
\nonumber\\
&&
\label{xpinLsav0}\\[2mm]
&=& \frac{32\pi\lambda^{n+1}}{c_F}\frac{(3\pi)^nL^2}{n}
\int\dd\beta\dd\gamma\,\, \delta(G)
\left[ \prod_{m=1}^n \sigma_m^3 t_m \right]
\,I_n(\beta,\gamma;L)
\label{xpinLsav}
\eea
The $\sigma_m$ that still appear in (\ref{xpinLsav0}) and (\ref{xpinLsav}) must now be viewed as
functions of the variables
of integration determined by the $n-1$ relations (\ref{recrhomit}) and 
sum rule (\ref{sumrulesigmatau}); we have furthermore
abbreviated
\beq
I_n(\beta,\gamma;L) = \int_{0}^\infty\dd\sav\,\sav^{3n-1}
\ee^{{\cal K}(\sav,\beta,\gamma;L)-\lambda\cV(\sav,\beta,\gamma;L)},
\label{dIbgL}
\eeq
in which we have expressed $\cV$ as a function of the new variables of
integration using the notation $\cV(\bS_1,\ldots,\bS_n;L) =\cV(\sav,\beta,\gamma;L)$,
and where we have shown explicitly the dependence of $\cK$ on the same variables.
The factor $\delta(G)/2\pi$ in (\ref{xpinLsav0}) and
(\ref{xpinLsav}) enforces the no-spiral constraint (\ref{nospiral}).
\vspace{3mm}

Up until this point all transformations of variables
applied to the initial expression (\ref{xpinL}) have been exact,
with the exception of the discussion in section \ref{secrange},
where an exponentially small contribution in $n$ was omitted.
Eq.\,(\ref{xpinLsav}) now expresses $\pi_n(L)$ entirely as an integral
over $2n$ angular variables and over the single radial variable $\sav$.
In our work \cite{Hilhorst05b} on the sidedness problem in 2D 
a radial integration occurred that could be done almost
trivially. Such is not the case here, and from this point on we will have
recourse to an expansion in inverse powers of $n$.

 
\section{Expansion in powers of $n$}
\label{seclargenexpansion}

Our approach will consist in finding, in the $2n$-dimensional phase space,
the maximum of the integrand on the RHS of Eq.\,(\ref{xpinLsav}), and to show
that an expansion about this maximum is possible.

\subsection{Scaling with $n$}
\label{secscalingwithn}

In order to carry out a large-$n$ expansion of 
$\pi_n(L)$ as given by equations (\ref{xpinLsav})-(\ref{dIbgL}),
we hypothesize for the various variables involved in the calculation
the following scaling with $n$,
\bea
r(\phi),s(\phi),s_m\,,\sav       &\sim& n^{1/3},    \nonumber\\[2mm]
                               L &\sim& n^{-1/6},   \nonumber\\[2mm] 
\tau_m\,,\,\beta_m\,,\, \gamma_m &\sim& n^{-1/2},   \nonumber\\[2mm]
                \xi_m\,,\,\eta_m &\sim& n^{-1},     
\label{scalingcV}
\eea
with $\xi_m$ and $\eta_m$ defined in the caption of figure \ref{figure_3}.
The scaling of $\sav$ and $L$ is suggested by recent work \cite{HilhorstLazar14}, whereas
the scaling of the angles $\beta_m$ and $\gamma_m$
is taken from reference \cite{Hilhorst05b}.
The latter implies the scaling of the $\tau_m$ {\it via\,} equations (\ref{recrhomit}) and (\ref{dtaum}).
The scaling of the $\xi_m$ and $\eta_m$ is a consequence of the sum rules
$\sum_{m=1}^n\xi_m=\sum_{m=1}^n\eta_m=0$.
Obviously, in each
of the two sums $\xi_m=\beta_m+\gamma_m$ and $\eta_m=\gamma_m+\eta_{m+1}$
the order $n^{-1/2}$ contributions must cancel.

We will encounter below many different
expressions containing sums of exactly $n$ or of ${\cal O}(n)$ terms,
where each term is a random variable depending on the $\beta_m$ and $\gamma_m$.
It was shown in detail in Ref.\,\cite{Hilhorst05b}
how the scaling with $n$ of such sums may be
determined. The basic rule is that the sum of $\sim n$ random variables
of zero average scales with an extra factor $n^{1/2}$
and the sum of variables of nonzero average with an extra factor $n$. 
This rule is complicated by the fact that the averages of $\beta_m$ and
$\gamma_m$ are $\sim n^{-1}$ but their rms deviations of order $n^{-1/2}$,
as well as by the occurrence of products of correlated variables
(such as $\tau_m\gamma_m$).
Our expansion in negative powers of $n$ will take all this into account,
and we will occasionally refer to Ref.\,\cite{Hilhorst05b} for details.

The fact, shown below, that the perturbation expansion leads to 
a series with finite coefficients will be considered by us as proof of the 
correctness of the assumed scaling (\ref{scalingcV}).




\subsection{Expansion of\, $\cV$}
\label{secexpansioncV}

We consider first the pumpkin volume $\cV$ whose expression is given by
(\ref{VsumV}). The function $g(x)$ occurring there is defined by (\ref{xgx})
and has the small $x$ expansion
\beq
g(x) = 1-\tfrac{1}{2}x^2 - \tfrac{1}{8}x^4 + {\cal O}(x^5),
\label{xgxsmall}
\eeq
which we will use in equation (\ref{VsumV}).
Equations (\ref{defxm}) and (\ref{scalingcV}) show that $x(\phi)$ scales as
$\sim n^{-1/2}$, so that (\ref{VsumV}) may be expanded as
\beq
\cV = \frac{1}{2\pi}\sum_{m=1}^n\int_{-\beta_m}^{\gamma_m}
\dd\phi\,\,2\pi^2 r^3(\phi)\left[\,1\,-\,\frac{L^2}{2r^2(\phi)}
\,-\, \frac{L^4}{8r^4(\phi)} + {\cal O}(n^{-3}) \right].
\label{expcV1}
\eeq
We will use the two expressions (\ref{relrphi}) for $r(\phi)$
and expand these in turn,
abbreviating $\epsilon_m=\cos^2\phi/\cos^2\beta_m-1$ so that
$\epsilon_m \sim L^2/s_{m+1}^2 \sim n^{-1}$. 
Expanding for large $n$ we get from the first equality of equation (\ref{relrphi})
\bea
r(\phi) &=& s_{m-1}\left[ 
1 + \frac{1}{2}\left\{ \epsilon_m+\frac{L^2}{s_{m-1}^2}\right\} 
  - \frac{1}{8}\left\{ \epsilon_m+\frac{L^2}{s_{m-1}^2}\right\}^2  
  +\ldots
\right] \nonumber\\[2mm]
&=& s_{m-1}\left[
1 +\frac{\epsilon_m}{2} +\frac{L^2}{2s_{m-1}^2}  
  -\frac{\epsilon_m L^2}{4s_{m-1}^2} - \frac{L^4}{8s_{m-1}^4}
  + \frac{c}{n^{2}} + {\cal O}(n^{-3}) 
\right],
\label{expRmin}
\eea
in which we have adopted a convention that we will 
use repeatedly below: the symbol ``$c$''
stands for an expression, each time a different one, 
that may depend on $\beta,\gamma,$ and $\sav$,
is of order $n^0$ as $n\to\infty$, but {\it does not depend on\,} $L$. 
In fact, inside the square brackets in equation (\ref{expRmin})
the second and third term are of order $n^{-1}$ and the fourth and fifth term
represent the full $L$ dependent contribution to order $n^{-2}$.

The second equality of equation (\ref{relrphi}) gives the same result as
(\ref{expRmin}) up to
the substitutions $s_{m-1}\mapsto s_m$ and $\beta_m\mapsto\gamma_m$. 
In each sector these two expressions are equivalent.
We will find it convenient to use below the first one for
$\phi>0$ and the second one for $\phi<0$.

After inserting the expansions (\ref{expRmin}) for $r(\phi)$ in (\ref{expcV1}),
also expanding $\epsilon_m$ to quadratic order in the angles,
doing the $\phi$ integrals, and rearranging terms, we get
\beq
\cV=\frac{1}{2\pi}\sum_{m=1}^n 2\pi^2 s_m^3
\left[ \gamma_m + \beta_{m+1} + \gamma_m^3 + \beta_{m+1}^3 +
\frac{L^2}{s_m^2}(\gamma_m + \beta_{m+1}) +\frac{c}{n^2}
+{\cal O}(n^{-3}) \right]
\label{expcV2}
\eeq
At order $n^{-1}$ we might have expected terms proportional to $L^4$  
in (\ref{expcV2}), but these appear to cancel 
After introducing the variables $\sav$ and $\tau_m$, and expanding for small
$\tau_m$,  there appear sums of products of the $\tau_m, \gamma_m$, and
$\beta_m$. 
We define the following expressions 
that are all of order $n^0$ as $n$ gets
large, 
\vspace{1mm}
\begin{equation}
\begin{array}{ll}
{\displaystyle \Gtwo(\beta,\gamma) = \frac{n}{2\pi}\sum_{m=1}^n \tau_m(\gamma_m+\beta_{m+1})},
& \\[2mm]
{\displaystyle \Ftwo(\beta,\gamma)  = \frac{1}{2}\sum_{m=1}^n (\gamma_m^2+\beta_{m+1}^2)}, 
&
{\displaystyle \tFtwo(\beta,\gamma) = \frac{n}{6\pi}\sum_{m=1}^n (\gamma_m^3+\beta_{m+1}^3)}, 
\\[2mm]
{\displaystyle \Ffour(\beta,\gamma) = \sum_{m=1}^n \tau_m^2},
&
{\displaystyle \tFfour(\beta,\gamma) = \frac{n}{2\pi}\sum_{m=1}^n \tau_m^2
(\gamma_m+\beta_{m+1})}. 
\end{array}
\label{array1}
\end{equation}
We recall here that the $\tau_m$ are functions of the $\beta_m$ and $\gamma_m$
defined implicitly by (\ref{dtaum}), (\ref{recrhomit}), (\ref{dG}), and (\ref{nospiral}).
Using that $\sum_{m=1}^{n}(\beta_{m}+\gamma_m)=2\pi$ we get
from (\ref{expcV2}) and (\ref{array1})
\beq
\cV(\sav,\beta,\gamma;L) = 2\pi^2\sav^3 \bigg[ \,1\,+\,\frac{L^2}{\sav^2} \,+\,\frac{W}{n}
\,+\,\frac{L^2}{\sav^2}\frac{\Uop}{n}\,+\,\frac{c}{n^2} 
\,+\,{\cal O}(n^{-3})\bigg]. 
\label{expcV3}
\eeq
in which 
\beq
W = 3\Gtwo+3\tFfour+3\tFtwo\,.
\label{dW}
\eeq
The prefactor $2\pi^2\sav^3$ in (\ref{expcV3}) is the volume of a torus whose 
major and minor radius are both equal to $\sav$
(sometimes called a {\it horn torus}).
With the scaling assumed in (\ref{scalingcV}) this prefactor is $\sim n$ as
$n\to\infty$. 
Inside the brackets in (\ref{expcV3}),
where $L^2/\sav^2\sim n^{-1}$, we have included 
all terms of order $n^{-1}$,
as well as the $L$ dependent term of order $n^{-2}$.
The $L$ independent terms of order $n^{-2}$ are indicated as $c/n^2$, where
$c$ is left undetermined.
At order $n^{-1}$ we might have expected terms 
proportional to $L^4/\sav^4$ and to
$(L^2/\sav^2)(\gamma_m^3+\beta_m^3)$, but both appear to cancel.


\subsection{Expansion of\, ${\cal K}$}
\label{secexpansionK}

We consider now the quantity $\ee^{\cK}$ that resulted from the integration
over the polarv angles and is given by (\ref{defcalK}) and (\ref{xKm}).
From Eq.\,(\ref{xKm}) we find by straightforward expansion
that
\beq
K(x)= 1 + \tfrac{1}{3}x^2 + {\cal O}(x^5)
    = \exp\big[ \tfrac{1}{3}x^2-\tfrac{1}{18}x^4 + {\cal O}(x^5) \big],
\label{expKx}
\eeq
which when substituted in (\ref{defcalK}) leads to 
\beq
{\cal K}(\sav,\beta,\gamma;L)
= \frac{nL^2}{3\sav^2} + \frac{L^2}{\sav^2}\Xop
- \frac{nL^4}{18\sav^4} + {\cal O}(n^{-2}) 
\label{explogK}
\eeq
Since $L^2/\sav^2 \sim n^{-1}$ we see that
the first term on the RHS of (\ref{explogK})
is of order $n^0$ and the second and third are of order $n^{-1}$.


\subsection{Expansion of ${\cal K} - \lambda{\cal V}$}
\label{secexpansionKcV}

Taking (\ref{expcV3}) and (\ref{explogK}) together we have
\bea
{\cal K}-\lambda{\cal V} &=& -2\pi^2\lambda\sav^3 \nonumber\\[2mm]
&& -2\pi^2\lambda\sav L^2 - 2\pi^2\lambda\sav^3 \frac{W}{n} 
+\frac{nL^2}{3\sav^2} \nonumber\\[2mm]
&& +\frac{L^2}{\sav^2}\Xop - \frac{nL^4}{18\sav^4}
-2\pi^2\lambda \sav L^2\frac{\Uop}{n} + \frac{c}{n} \nonumber\\[2mm]
&& +\,{\cal O}(n^{-2}),
\label{xKmV}
\eea
where we have arranged terms such that the ones in the first, second, and
third line are 
${\cal O}(n)$, ${\cal O}(n^0)$, and ${\cal O}(n^{-1})$, respectively.
To see this it suffices to know that the only hidden factors of $n$ are those
in $\sav^3\sim n$ and $L^2/\sav^2\sim n^{-1}$. 


\section{Integrating over the face radius $\sav$}
\label{secintegratingsav}


\subsection{The integral $I_n(\beta,\gamma;L)$}
\label{secintegral}

We substitute result (\ref{xKmV}) for ${\cal K}-\lambda{\cal V}$
in the integral $I_n(\beta,\gamma;L)$ defined by (\ref{dIbgL}).
Let dimensionless scaled variables $x$ and $\Lambda$ be defined by
\bea
      x &=& (2\pi^2\lambda)^{1/3}n^{-1/3}\sav, \nonumber\\[2mm] 
\Lambda &=& (2\pi^2\lambda)^{1/3}n^{ 1/6}L.
\label{scaledvar}
\eea
In terms of these integral (\ref{dIbgL}) may be written
\beq
I_n = \left( \frac{n}{2\pi^2\lambda} \right)^n
\int_0^\infty\dd x\,\,\ee^{f(x)}\left[ 1+\frac{\Aop(x)}{n}+\frac{c}{n}
+ {\cal O}(n^{-2}) \right],
\label{dI3}
\eeq
in which $\Aop(x)/n$ stands for the sum of the first three terms in the third
line of (\ref{xKmV}), that is, in terms of $x$ and $\Lambda$,
\beq
\Aop(x) = \frac{\Lambda^2}{x^2}F_4 - \frac{\Lambda^4}{18x^4} -x\Lambda_2 G_2\,,
\label{dY}
\eeq
and where we define
\beq
f(x) = (3n-1)\log x - nx^3 - x^3W - \Lambda^2\left(x-\frac{1}{3x^3}\right).
\label{deffx}
\eeq
For $L=0$, hence $\Lambda=0$, the integral is easy to calculate in closed form.
For the general case $L\geq 0$
we will calculate $I_n$ in the limit of large $n$,
with $L\sim n^{-1/6}$, as stated in our hypothesis (\ref{scalingcV}).


\subsection{Saddle point expansion}
\label{secsaddle}

We will carry out the integral (\ref{dI3}) by means of a saddle point
expansion.
We calculate the leading order term and the corrections of relative order
$n^{-1}$, limited to those that are dependent on $L$.

The saddle point condition $\dd f(x)/\dd x=0$ applied to (\ref{deffx})
has the solution $x=x_*$ where
\beq
x_* = 1 + \frac{x_1}{n} + \frac{x_2}{n^2} + {\cal O}(n^{-3})
\label{xstar}
\eeq
in which 
\beq
x_1=-\frac{1}{9}-\frac{W}{3}-\frac{5\Lambda^2}{27}\,.
\label{xx1}
\eeq
The explicit expression for $x_2$ is easy to find but
will drop out of later calculations. 
Setting $f_*\equiv f(x_*)$ we find by substitution of (\ref{xstar}) in 
(\ref{deffx}) that
\beq
f_* = -n-W-\frac{2}{3}\Lambda^2 +\frac{9}{2n} x_1^2 + {\cal O}(n^{-2}).
\label{fstar}
\eeq
This, combined with (\ref{xx1}), in turn leads to the expansion
\beq
\ee^{f_*} = {\ee^{-n}}\,\ee^{-W-\frac{2}{3}\Lambda^2}
\left[ 1 + \frac{1}{n}\left(\frac{5}{27}(1+3W)\Lambda^2 
+ \frac{25}{162} \Lambda^4 \right) + \frac{c}{n} + {\cal O}(n^{-2})\right],
\label{expfstar}
\eeq
where only the $\Lambda$ dependent terms in $x_1^2$ have been included
explicitly, the remaining ones being absorbed by $c$.
We recall here our convention to let $c$ stand for terms that are
of order $n^0$ but do not depend on $\Lambda$, whereas ${\cal O}(n^{-k})$
indicates any terms, whether $\Lambda$ dependent or not, that are of order
$n^{-k}$. For the $k$th derivative $f_*^{(k)}$ at the saddle point we have
\bea
f_*^{(2)} &=& -9n+1-6W+2\Lambda^2 + {\cal O}(n^{-1}), \nonumber\\[2mm]
f_*^{(3)} &=& {\cal O}(n^0), \nonumber\\[2mm]
f_*^{(4)} &=& -18n + {\cal O}(n^0). 
\label{fstarderiv}
\eea
The $f_*^{(k)}$ with $k\geq 5$ are all of order $n$ but will not be needed
in the calculation.
In (\ref{dI3}) we pass to the new variable of integration
$u\equiv x-x_*$ and Taylor expand $f(x)$ and $\Aop(x)$ about $x=x_*$ 
in powers of $u$.
\bea
I_n &=& \left( \frac{n}{2\pi^2\lambda} \right)^n
\ee^{f_*} \left[ 1+\frac{\Aop(1)}{n} + \frac{c}{n} +{\cal O}(n^{-2}) \right]
\nonumber\\[2mm]
&& \times\int_{-\infty}^{\infty} \dd u\,\,\ee^{\frac{1}{2} u^2 f_*^{(2)}}
\left[ 1 + \frac{1}{24}u^4f_*^{(4)} + {\cal O}(n^{-2}) \right]
\label{dI4}
\eea 
The contributions of
terms with odd order derivatives under the integral sign
vanish by symmetry.
Since $f_*^{(2)}$ scales as $\sim n$,
it follows that $u$ scales as $n^{-1/2}$ and,
in view of (\ref{fstarderiv}), that $u^4f_*^{(4)}$ scales as $\sim n^{-1}$.
Upon carrying out the $u$ integration in (\ref{dI4}) we obtain 
\bea
I_n &=& \left( \frac{n}{2\pi^2\lambda} \right)^n 
\left[ 1+\frac{\Aop(1)}{n} 
+ \frac{c}{n} + {\cal O}(n^{-2}) \right]
\nonumber\\[2mm]
&&\times\, \sqrt{\frac{2\pi}{\big| f_*^{(2)} \big|}}\,\,
\exp(f_*)
\left[ 1+\frac{f_*^{(4)}}{8\big[ f_*^{(2)} \big]^2} 
+ {\cal O}(n^{-2}) \right].
\label{dI5}
\eea
We have that 
$f_*^{(4)}/\big( 8\big| f_*^{(2)} \big|^2 \big) = -1/(36n) + {\cal O}(n^{-2})$,
which to leading order is independent of $\Lambda$ and may therefore be
absorbed in the term $c/n$.
After expanding the square root in (\ref{dI5}),
\beq
\big| f_*^{(2)} \big|^{-1/2} 
= \frac{1}{3n^{1/2}}\left[ 1 
+ \frac{\Lambda^2}{9n} +\frac{c}{n} +{\cal O}(n^{-2})
\right],
\label{expfdp}
\eeq
and using Stirling's formula
$(n/\ee)^n\sqrt{2\pi n} = n!\,[1+c/n+{\cal O}(n^{-2})]$ we get 
\beq
I_n(\beta,\gamma;L) = \frac{(n-1)!}{3(2\pi^2\lambda)^n}\,
\ee^{-W-\frac{2}{3}\Lambda^2}
\left[ 1 + \frac{1}{n} \left( \Atwo\Lambda^2 + \Afour\Lambda^4 + c \right) 
+ {\cal O}(n^{-2}) \right]
\label{resIfinal}
\eeq
in which 
\beq
\Atwo = \frac{8}{27} + \frac{2}{3}\Gtwo + \frac{5}{3}\Ftwo + 
\frac{5}{3}\tFfour + \Ffour\,, \qquad
\Afour = \frac{8}{81}\,,
\label{xAtwofour}
\eeq
and where $c$ is a quantity of order $n^0$ and independent of $\Lambda$
that we may leave undetermined.

Equations (\ref{resIfinal}) and (\ref{xAtwofour}) 
complete the calculation of the integral $I_n$.
The $\beta$ and $\gamma$ dependence of the result is contained in 
the quantities $W, \Atwo, \Afour$, given by (\ref{dW}) and (\ref{array1}), 
and in $c$.


\subsection{Average face radius and approach to a circle} 
\label{seccellradius}

For given edgedness $n$
the integration over the face radius $\sav$, defined by (\ref{dravrhom}),
has a maximum at $\sav = \sfS_n$.
From (\ref{scaledvar}) and (\ref{xstar}) we see that we have, 
to leading order in $n$,
\beq
\sfS_n \simeq (2\pi^2\lambda)^{-1/3}n^{1/3}.
\label{xsfSn}
\eeq
Near the maximum the integrand is a Gaussian of width
\bea
\la (\sav-\sfS_n)^2 \ra &=& (2\pi^2\lambda)^{-2/3}n^{2/3}\la (x-x^*)^2\ra
\nonumber\\[2mm]
&=&(2\pi^2\lambda)^{-2/3}n^{2/3}/\big| f_*^{(2)} \big|
\nonumber\\[2mm]
&=& \frac{1}{9}(2\pi^2\lambda)^{-2/3}n^{-1/3}.
\label{xvarsfSn}
\eea
Since $\la (\sav-\sfS_n)^2 \ra^{1/2} / \sfS_n \simeq 1/(3n^{1/2})$, 
the fluctuations about $\sfS_n$ are negligible in the large-$n$
limit and therefore $\sfS_n$ is also the {\it average\,} face radius.
\vspace{2mm}

We must now investigate the fluctuations of an
individual variable $s_m$, denoting the distance
between the center of the face and its $m$th edge, 
and the radius $\sav$, which is the average of all $n$ such distances
[see equation (\ref{dravrhom})].
We find by combining previous results that
\beq
\frac{1}{n}\sum_{m=1}^n\langle(s_m-\sav)^2\rangle =
\frac{1}{n}\sum_{m=1}^n\langle \sav^2\tau_m^2 \rangle,
\label{circle1}
\eeq
where we used (\ref{dravrhom}) and (\ref{dtaum}).
Now, knowing that $\sav$ is strongly peaked around its average $\sfS_n$, we
may, to leading order, take it out of the angular brackets in (\ref{circle1}).
This leads to 
\beq
\frac{1}{n}\sum_{m=1}^n\langle(s_m-\sav)^2\rangle \,=\,
\frac{\sfS_n^2}{n}\sum_{m=1}^n\langle\tau_m^2\rangle
\,=\, f_4 \frac{\sfS_n^2}{n}
\sim n^{-1/3}.
\label{circle2}
\eeq
This shows that for $n\to\infty$ the individual $s_m$ all get infinitely
sharply peaked around the average $\sfS_n$, and we may extend (\ref{xsfSn}) to 
\beq
\la s_m\ra \simeq \sfS_n \simeq (2\pi^2\lambda)^{-1/3}n^{1/3},
\qquad m=1,2,\ldots,n.
\label{xsfSnext}
\eeq 
This last equation implies that the shape of the interface tends to a circle
of radius $\sfS_n$ as given by (\ref{xsfSn}).


\section{Transforming the angular averages}
\label{sectransforming}

Upon substituting equation (\ref{resIfinal}) in (\ref{xpinLsav}) 
and using (\ref{scaledvar}) for $\Lambda$ we obtain
\bea
\pi_n(L) &=& \frac{32\pi}{3c_F}\frac{(n-1)!}{n} 
\left( \frac{3}{2\pi} \right)^n \lambda L^2
\int\dd\beta\dd\gamma\,\delta(G) \nonumber\\[2mm]
&& \times \left[ \prod_{m=1}^n \sigma_m^3t_m \right]
\ee^{-W-\frac{2}{3}\Lambda^2}
\left[ 1 + \frac{{\cal R}_\Lambda}{n}
+ {\cal O}(n^{-2}) \right], 
\label{xpinLI}
\eea
with the symbol $\int\dd\beta\dd\gamma$ defined in (\ref{defintbetagamma})
and where we have abbreviated
\beq
{\cal R}_\Lambda = \Atwo\Lambda^2+\Afour\Lambda^4+c.
\label{defcalR}
\eeq
The following development closely parallels the one for the 2D Voronoi 
cell that was carried out in Ref.\,\cite{Hilhorst05b} 
(see also Ref.\,\cite{Hilhorst07}, Appendices A and B).
Our description will therefore be succinct.

Rather than using the set of variables $\{\beta_m,\gamma_m|m=1,\ldots,n\}$
we will employ the sets $\xi\equiv \{\xi_m|m=1,\ldots,n\}$
and $\eta\equiv \{\eta_m|m=1,\ldots,n\}$; these variables have been defined in
the caption of figure \ref{figure_3}.
Inversely, the $\beta_m$ and $\gamma_m$ may be
expressed in terms of the sets $\xi$ and $\eta$ {\it and\,} one of the 
$\beta_m$, let us say $\beta_1$. We have
\bea
\beta_m &=& \phantom{-}\beta_1 - 
\sum_{\ell=1}^{m-1}(\xi_{\ell}-\eta_\ell),
\nonumber\\
\gamma_m &=& -\beta_1 +
\sum_{\ell=1}^{m-1}(\xi_{\ell}-\eta_{\ell}) + \xi_m\,, 
\qquad m=1,\ldots,n.
\label{inversebgxy}
\eea
It would seem that $\beta_1$ 
need be given. However,
it was shown in Ref.\,\cite{Hilhorst07} that the
no-spiral constraint $G=0$ of Eq.\,(\ref{nospiral}) above, 
when rewritten with the aid of
(\ref{inversebgxy}) in terms of $\xi, \eta$, and $\beta_1$, 
has a unique solution $\beta_1=\beta_*(\xi,\eta)$. We 
write
\beq
\delta(G)=\frac{\delta(\beta_1-\beta_*)}{G'}
\label{deltabetastar}
\eeq
in which $G'\equiv\dd G(\xi,\eta;\beta_1)/\dd\beta_1$ where the
derivative is taken at fixed $\xi,\eta$. 

The $\xi_m$ and $\eta_m$ are necessarily nonnegative.
We abbreviate the integration on them as
\beq
\int\dd\xi\dd\eta = \int_0^\infty\left[ \prod_{m=1}^n\dd\xi_m
\dd\eta_m \right]
\,\delta\Big( \sum_{m=1}^n\xi_m-2\pi \Big)
\delta\Big( \sum_{m=1}^n\eta_m-2\pi \Big).
\label{defintxieta}
\eeq
The presence of the two delta functions in definition
(\ref{defintxieta}) 
has allowed us to take the upper limits of the integrations
equal to infinity.
We furthermore need
conditions on $\xi,\eta$ that will guarantee that
$\beta_m,\gamma_m<\frac{\pi}{2}$.
We will represent these conditions by the indicator function
\beq
\Theta(\xi,\eta) = \prod_{m=1}^n \left[
\theta\Big( \frac{\pi}{2}-\beta_m \Big)
\theta\Big( \frac{\pi}{2}-\gamma_m \Big) \right].
\label{defTheta}
\eeq
The change of variables of integration in (\ref{xpinLI}) may then be 
written as
\beq
\int\dd\beta\dd\gamma \,\,\delta(G) =
\int\dd\xi\dd\eta \,\,\frac{1}{G'}\Theta(\xi,\eta).
\eeq
With the additional definition
\beq
\ee^{-{\mathbb{V}}} = \frac{1}{G'} \prod_{m=1}^n\frac{\sigma_m^3 t_m}{\xi_m}
\,\,\ee^{-W}
\label{dmatV}
\eeq
we may then rewrite (\ref{xpinLI}) as
\bea
\pi_n(L) &=& \frac{32\pi}{3c_F}\frac{(n-1)!}{n} 
\left( \frac{3}{2\pi} \right)^n \lambda L^2\,\ee^{-\frac{2}{3}\Lambda^2}
\int\dd\xi\dd\eta
\,\,\xi_1\xi_2\ldots\xi_n\nonumber\\[2mm]
&& \times\,
\Theta\,\ee^{-\mathbb{V}}
\Big[ 1 + \frac{{\cal R}_\Lambda}{n}
+ {\cal O}(n^{-2}) \Big].
\label{xpinLxieta}
\eea
Since \cite{Hilhorst07}
\beq
{\cal N} \equiv \int\dd\xi\dd\eta\,\, \xi_1\xi_2\ldots\xi_n
= \frac{(2\pi)^{3n-2}}{(2n-1)!(n-1)!}\,,
\label{normalization}
\eeq
we may multiply this quantity into the prefactor on the RHS of 
(\ref{xpinLxieta})
and rewrite that relation in the compact form
\beq
\pi_n(L) = \frac{32\pi}{3c_F}\times\frac{2(12\pi^2)^{n-1}}{(2n)!}
\times\lambda L^2\,\ee^{-\frac{2}{3}\Lambda^2}
\Big\langle 
\Theta\,\ee^{-{\mathbb V}}
\big[ 1 + \frac{{\cal R}_\Lambda}{n}
+ {\cal O}(n^{-2}) \big] 
\Big\rangle_{\:0},
\label{xpinLav0}
\eeq
in which for any $\cA$ the average $\la\cA\ranul$ is defined as 
\beq
\la\cA\ranul = \frac{1}{{\cal N}}\int\dd\xi\dd\eta\,\xi_1\xi_2\ldots\xi_n\,\cA.
\label{defav0}
\eeq
We now consider $\Theta$.
For large $n$ the angles will all become small:
$\xi_m, \eta_m\sim n^{-1}$ and $\beta_m,\gamma_m\sim n^{-1/2}$,
and the conditions imposed by $\Theta$ are violated with a probability
that is exponentially small in $n$.
In our expansion in powers of $n$ we may therefore set $\Theta=1$. 
This leads us to rewrite (\ref{xpinLxieta}) as the final result of this
section, 
\beq
\pi_n(L) = \frac{32\pi}{3c_F}\times\frac{2(12\pi^2)^{n-1}}{(2n)!}
\times\lambda L^2\,\ee^{-\frac{2}{3}\Lambda^2}
\big\langle \ee^{-\mathbb{V}} \big\rangle_{_0}
\big[ 1+\frac{\big\langle {\cal R}_\Lambda \big\rangle}{n} 
+ {\cal{O}}(n^{-2})\big],
\label{xpinLav1}
\eeq
in which for any $\cA$ the average $\la\cA\ra$ is defined as 
\beq
\la\cA\ra = \frac{\la\cA\ee^{-\mathbb{V}}\ranul}
{\la \ee^{-\mathbb{V}}\ranul}\,,
\label{defav1}
\eeq
and where $\mathbb{V}$ and ${\cal{R}}_\Lambda$ are given by (\ref{dmatV})
and (\ref{defcalR}), respectively.
At this point we may notice that to leading order, that is, in the absence of the term ${\cal R}_\Lambda/n$, the $\Lambda$ (or: $L$) dependence of $\pi_n(L)$ 
has been factorized out. 
\vspace{2mm}

Our initial problem (\ref{xpinLxi}) was to evaluate $\pi_n(L)$
an integral on $n$ first-neighbor
positions $\bR_1,\ldots,\bR_n$, that is, on $3n$ variables.
After we carried out in section \ref{secintegratingtheta}
the integrals over the polar angles $\theta_1,\ldots,\theta_n$
there remained $2n$ variables of integration.
Subsequent to  further transformations and a large $n$ expansion
we have, in equation (\ref{xpinLav1}), arrived at a $2n$-fold integration 
on the variables $\xi$ and $\eta$, represented by the angular brackets $\langle\ldots\rangle$ and $\langle\ldots\rangle_0$. 

The highly nontrivial fact about equation (\ref{xpinLav1}) is that 
$\big\langle \ee^{-\mathbb{V}} \big\rangle_{_0}$
tends to a constant for $n\to\infty$. We will show this and determine the
value of that constant in the next section. 


\section{Calculation of  $\big\langle \ee^{-\mathbb{V}} \bigranul$ }
\label{seccalculationexpV}

It will appear that apart from a change of coefficients the calculation of
$\big\langle \ee^{-\mathbb{V}} \big\rangle_{_0}$,
with $\ee^{-\mathbb{V}}$ defined by (\ref{dmatV}),
is identical to those that we performed for the
two-dimensional Voronoi cell \cite{Hilhorst05b} and for a family of line
tessellations \cite{HilhorstCalka08}.
We will therefore heavily rely here on this earlier work. 

In (\ref{dmatV}) the factor multiplying
$\ee^{-W}$ may be shown by the methods of Ref.\,\cite{Hilhorst05b} to be equal to
\beq
\frac{1}{G'} \prod_{m=1}^n \frac{\sigma_m^3 t_m}{\xi_m}
= \exp\Big( \!-\Ftwo+\frac{3}{2}\Ffour \Big)
\Big[ 1 + {\cal O}(n^{-1/2}) \Big],
\label{factor}
\eeq
with $\Ftwo$ and $\Ffour$ given in (\ref{array1}) and 
where $L$-independent correction terms of order $n^{-1/2}$ appear 

\beq
\tFtwo  = \Ftwo  + {\cal O}(n^{-1/2}),
\qquad
\tFfour = \Ffour + {\cal O}(n^{-1/2}). 
\label{elimtilde}
\eeq
Upon combining (\ref{factor}) and (\ref{elimtilde}) with (\ref{dmatV}) and (\ref{dW})
we find that
\beq
{\mathbb{V}} = \sfA\Ftwo + 2\sfB\Ffour + 2\sfC\Gtwo + {\cal O}(n^{-1/2})
\label{xmatV}
\eeq
with
\beq
\sfA=2, \qquad \sfB=\frac{9}{4}\,, \qquad \sfC=\frac{3}{2}\,.
\label{xsfABC}
\eeq
For the 2D Voronoi cell \cite{Hilhorst05b} we had at this stage a similar
expression but with $\sfA=\sfB=\sfC=1$,
and for the line tessellation problem \cite{HilhorstCalka08,HugSchneider07} we had 
a family of expressions such that
$\sfA=\alpha-1$, $\sfB=\frac{1}{4}\alpha^2$, and $\sfC=\frac{1}{2}\alpha$.
Our present problem, equation (\ref{xsfABC}), 
belongs to the same family and has $\alpha=3$. 

We may discuss without extra effort, from here up to and including
equation (\ref{rC30}), the general expression (\ref{xmatV})
with three arbitrary constants $\sfA,\sfB$, and $\sfC$.  
%
The expressions for $\Gtwo, \Ftwo$, and $\Ffour$ have been defined in
(\ref{array1}) in terms of the variables $\beta_m$ and $\gamma_m$, 
but may be expressed in terms of the
$\xi_m$ and $\eta_m$ with the aid of (\ref{inversebgxy}) and the known value
$\beta_1=\beta_*(\xi,\eta)$. 
Knowing that $\la\xi_m\ranul=\la\eta_m\ranul=2\pi n^{-1}$ 
we define the scaled deviations from average
\beq
\delta x_m = n(\xi_m -2\pi n^{-1}), \qquad
\delta y_m = n(\eta_m-2\pi n^{-1}),
\label{xdeltaxy}
\eeq
which vary on scale $n^0$.
Let their Fourier transforms be 
\beq
\hat{X}_q= \frac{1}{2\pi n^\half}\summ \ee^{2\pi{\rm i} qm/n} \delta x_m\,,
\qquad 
\hat{Y}_q= \frac{1}{2\pi n^\half}\summ \ee^{2\pi{\rm i} qm/n} \delta y_m\,, 
\label{FTdeltaxy}
\eeq
where $q=0,\pm 1,\pm 2,\ldots,\pm(\frac{1}{2}n-\frac{1}{2})$ 
for $n$ odd and 
$q=0,\pm 1,\pm 2,\ldots,\pm(\frac{1}{2}n-1),\frac{1}{2}n$ 
for $n$ even.
In terms of these we find from (\ref{array1}) that
\bea
F_k  &=&  \sum_{q\neq 0} \frac{1}{q^k}(\hat{X}_q-\hat{Y}_q)
(\hat{X}_{-q}-\hat{Y}_{-q})+{\cal O}(n^{-1/2}),
\nonumber\\[2mm]
%
\Gtwo  &=& \sum_{q\neq 0} \frac{1}{2q^2}
\big( \hat{X}_q\hat{Y}_{-q}+\hat{X}_{-q}\hat{Y}_q 
- 2\hat{Y}_q\hat{Y}_{-q} \big)+{\cal O}(n^{-1/2}).
\label{xFTFG}
\eea
It is useful to set ${\bsZ}_q=(\hat{X}_q,\hat{Y}_q)$.
Using (\ref{xFTFG}) and (\ref{elimtilde}) 
we may then write (\ref{xmatV}) as
\beq
{\mathbb V}=\sum_{q\neq 0}\, 
{\bsZ}_q\cdottt{\bsV}_q\cdottt{\bsZ}_{-q}^{\rm T}+{\cal O}(n^{-1/2}),
\label{V0compact}
\eeq
where {\sc T} stands for transposition and where ${\bsV}_q$ is the symmetric
matrix 
\beq
{\bsV}_q = \left(
\begin{array}{rr}
\sfA q^{-2}+2\sfB q^{-4}\phantom{xxxxx} & 
-(\sfA-\sfC)q^{-2}-2\sfB q^{-4} \\[2mm]
-(\sfA-\sfC)q^{-2}-2\sfB q^{-4} \phantom{xx}& 
(\sfA-2\sfC)q^{-2}+2\sfB q^{-4}
\end{array}
\right).
\label{exprBq}
\eeq
Let {\bf 1}$=\mbox{diag}\{1,1\}$ and $\bsE=\mbox{diag}\{1,2\}$. 
We define 
\bea
\Lambda_q &=& \det(\bse+\bsV_q\bsE) \nonumber\\[2mm]
          &=& 1\,+\,\frac{3\sfA-4\sfC}{q^2} + \frac{6\sfB-2\sfC^2}{q^4}\,.
\label{dLambdaq}
\eea
For the general $\mathbb{V}$ given by (\ref{V0compact}) and (\ref{exprBq}),
and using definition (\ref{defav0}) of the average $\la\ldots\ra_0$, we 
may the show by the methods of Ref.\,\cite{Hilhorst05b} that  
\beq
\big\langle \ee^{-\mathbb{V}} \bigranul
= \prod_{q=1}^\infty \Lambda_q^{-1} + {\cal O}(n^{-1/2}), \qquad n\to\infty.
\label{rC30}
\eeq
It was certainly not {\it a priori\,} evident
that this quantity is a finite constant in the limit $n\to\infty$.
The first result of this kind \cite{Hilhorst05b} was derived 
for the 2D Poisson-Voronoi cell.

We return now to the special values (\ref{xsfABC}) of $\sfA$, $\sfB$, and
$\sfC$ relevant to our problem. For these we define
$C(3)\equiv\lim_{n\to\infty}\big\langle \ee^{-\mathbb{V}} \bigranul$, 
and since in this special case $\Lambda_q=1+9q^{-4}$ we have
from (\ref{rC30})
\beq
C(3) = \prod_{q=1}^\infty \frac{q^4}{q^4+9} = 0.053891.
\label{rC3}
\eeq
This product on the wavenumbers $q$ has the interpretation of a partition
function, namely the one of the ``elastic'' deformations of the $n$-edged
face with respect to a circle, the elasticity being, of course, of purely entropic origin. 


\section{Calculation of $\la{\cal R}_\Lambda\ra$}
\label{seccalculationcR}

The work that remains to be done is the calculation of the 
coefficient $\la{\cal R}_\Lambda\ra$ in (\ref{xpinLav1}).
Obviously it suffices to find its limiting value as $n\to\infty$.

Although the variables of integration are not Gaussian, 
it was shown in Ref.\,\cite{Hilhorst05b}
that {\it to leading order in the large-$n$ expansion\,}
averages of type (\ref{defav0}) may be carried out
as if the $\xi_m$ and $\eta_m$ were distributed with the Gaussian weights
\begin{align}
\frac{1}{{\cal N}}\int\dd\xi\dd\eta\,\, \xi_1\xi_2\ldots\xi_n\,\cA
=& \nonumber\\[2mm]
\int\prod_q\dd\hat{X}_q\dd\hat{Y}_q\,&
\exp\Big(\! -\sumqno\hat{X}_q\hat{X}_{-q} 
          -\tfrac{1}{2}\sumqno\hat{Y}_q\hat{Y}_{-q} \Big)\,\cA.
+ {\cal O}(n^{-1/2})
\label{Gaussianweight}
\end{align}
The variances of the $\hat{X}_q$ and the $\hat{Y}_q$ in (\ref{Gaussianweight}) 
differ by a factor of two due to the appearance
of the product $\xi_1\xi_2\ldots\xi_n$ in the integration on the LHS.
The delta function constraints present in the definition (\ref{defintxieta})
of $\int\dd\xi\dd\eta$
have been incorporated in (\ref{Gaussianweight}): this Gaussian weight does not
depend on $\hat{X}_0$ and $\hat{Y}_0$, and neither should the otherwise
arbitrary integrand $\cA$. 

It was shown in Ref.\,\cite{Hilhorst05b} how correlations between the
$\hat{X}_q$ and $\hat{Y}_q$ may be calculated to leading order in the 
large-$n$ expansion. 
For ${\mathbb V}$ given by the general expression (\ref{V0compact})-(\ref{exprBq}) 
we obtain by the same method the basic correlations
\bea
\la\hat{X}_q\hat{X}_{-q}\ra &=& 
\frac{1}{2} -\frac{1}{2\Lambda_q} \Big[
\frac{\sfA}{q^2}+\frac{2(\sfB-\sfC^2)}{q^4} \Big] 
+{\cal O}(n^{-1/2}), \non
\la\hat{Y}_q\hat{Y}_{-q}\ra &=& 
1 - \frac{2}{\Lambda_q} \Big[
\frac{\sfA-2\sfC}{q^2}+\frac{2\sfB-\sfC^2}{q^4} \Big]
+{\cal O}(n^{-1/2}), \non
\tfrac{1}{2}\big[ \la\hat{X}_q\hat{Y}_{-q}\ra 
+ \la\hat{X}_{-q}\hat{Y}_{q}\ra \big]
&=& 
\frac{1}{\Lambda_q} \Big[\frac{\sfA-\sfC}{q^2}+\frac{2\sfB}{q^4} \Big]
+{\cal O}(n^{-1/2}).
\label{xbasiccorr}
\eea
We need the average $\la{\cal R}_\Lambda\ra$ with ${\cal R}_\Lambda$ given by 
(\ref{defcalR}), (\ref{xAtwofour}), and (\ref{xFTFG}).
Let us set 
\beq
f_k=\lim_{n\to\infty}\la F_k \ra \quad k=2,4, \qquad
g_2=\lim_{n\to\infty}\la G_2 \ra.
\label{deffg}
\eeq
After Fourier transforming the expressions for
$F_2, F_4$, and $G_2$, then using the basic correlations (\ref{xbasiccorr}),
and finally substituting our particular values (\ref{xsfABC}) of $\sfA,\sfB$,
and $\sfC$ we obtain 
\beq
f_2 = \phantom{-}\frac{3}{2}\sum_{q=1}^\infty\frac{q^2}{9+q^4} = 0.96119,
\qquad 
f_4 = \phantom{-}\frac{3}{2}\sum_{q=1}^\infty\frac{1}{9+q^4} = 0.23764,  
\label{xf2f4}
\eeq
and $g_2=-\frac{2}{3}f_2-f_4$. 
If we set $a_2=\la A_2\ra$, $a_4=A_4$, and $c_0=\la c\ra$, then
Eq.\,(\ref{defcalR}) becomes
\beq
\la{\cal R}_\Lambda\ra = a_2\Lambda^2+a_4\Lambda^4+c_0
\label{xavcR}
\eeq
in which, after elimination of $g_2$ and with the aid of (\ref{xf2f4}),
\beq
a_2 = \frac{8}{27}+\frac{11}{9}f_2+2f_4 = 1.94635\,, 
\qquad a_4 = \frac{8}{81} = 0.09877\,,
\label{xa2}
\eeq
and where $c_0$ has not been calculated.
Substituting (\ref{xavcR}) in (\ref{xpinLav1}) yields
\bea
\pi_n(L) &=& \frac{32\pi}{3c_F}\,\frac{2(12\pi^2)^{n-1}}{(2n)!}C(3)
\,\lambda L^2\,\ee^{-\frac{2}{3}\Lambda^2}
\nonumber\\[2mm]
&&\times \Big[ 1+\frac{c_0^\prime}{n^{1/2}}\frac{1}{n}\Big( a_2\Lambda^2+a_4\Lambda^4
+c_0 +{\cal O}(n^{-2})\Big) \Big],
\label{fpinL}
\eea
in which $c_0^\prime$ is an unknown but $\Lambda$-independent coefficient.
We suspect that in fact $c_0^\prime=0$, mainly because
in the related 2D sidedness problem numerical evidence \cite{Hilhorst07}
convincingly shows the
absence of correction terms of relative order $n^{-1/2}$. 
Eq.\,(\ref{fpinL}) is close to our final result.


\section{Final results for $p_n$ and $Q_n(L)$}
\label{secarriving}

We are now able to list our principal results. We factorize  $\pi_n(L)$ according to 
\beq
\pi_n(L) = Q_n(L)p_n\,,
\label{dQnL}
\eeq
where $p_n$ is the probability for the face to have $n$ edges and $Q_n(L)$ is
the conditional probability distribution of $L$ for given $n$.
Let us define the dimensionless scaling variable $y$ by
\bea
y=\kap n^{1/6}L, \qquad \kap &=& 2^{-1/6}3^{-1/2}\pi^{7/6}\lambda^{1/3}
\nonumber\\[2mm]
&=& 1.95558\,\lambda^{1/3},
\label{dy}
\eea
in which the choice of the inverse length constant $\kap$ will become clear below.
We may eliminate $\Lambda$ and $L$ from (\ref{fpinL}) 
in favor of $y$ using (\ref{dy})
and the relation $\Lambda = (6/\pi)^{1/2}y$ which follows from
(\ref{dy}) and (\ref{scaledvar}).
This leads to
\beq
p_n = \frac{1}{c_F}\left( \frac{3}{2\pi^5 n} \right)^{1/2}
\left[ \frac{(12\pi^2)^n}{(2n)!}C(3) \right]
\Big[ 1 + \frac{c_0^\prime}{n^{-1/2}} + \frac{c_0}{n} + {\cal O}(n^{-2}) \Big]
\label{fpn}
\eeq
and
\bea
Q_n(L)\dd L &=& {\cQ}_n(y)\dd y \nonumber\\[2mm]
&=& {\cQ}(y)
\Big[ 1 + \frac{1}{n} 
\Big( q_2 y^2 +q_4 y^4 - q_0 \Big) + {\cal O}(n^{-2})  
\Big] \dd y,
\label{fQnL}
\eea
in which the first line defines ${\cQ}_n(y)$,
where ${\cQ}(y)$ stands for the probability distribution
\beq
{\cQ}(y) = \frac{32 y^2}{\pi^2}\exp\left( -\frac{4y^2}{\pi} \right),
\qquad y>0,
\label{xQy}
\eeq
and where
\beq
q_2=\frac{6a_2}{\pi} = 3.71726\,, \qquad 
q_4 = \frac{32}{9\pi^2} = 0.36025.
\label{xqk}
\eeq
The coefficient $q_0$ in (\ref{fQnL}) is determined by the normalization
condition imposed on $Q_n(L)$ to order $n^{-1}$, which
leads to a redefinition of the unknown constant $c_0$ in (\ref{fpn}). 
Using that $\overline{y^2}=3\pi/8$ and $\overline{y^4}=15\pi^2/64$ we get
\beq
q_0 = \frac{9}{4}a_2 + \frac{5}{6} = 5.21263.
\label{xq0}
\eeq

The definition of $\kap$ in (\ref{dy}) is such that
$\overline{y}=1$, where the overbar denotes the average with respect to
${\cQ}(y)$. 
The higher moments, some of which are needed below, are given by
$\overline{y^k}=(\pi/4)^{(k-1)/2}\Gamma\big( (k+3)/2 \big)$.
\vspace{2mm}

The important point is that $L$ scales as $\sim n^{-1/6}$ and that its
probability distribution is fully known, including the leading
correction-to-scaling term. 
We now recall equations (\ref{xvarsfSn}) and (\ref{circle2}), which say
that this scale $n^{-1/6}$ is also the scale of the root-mean-square 
{\it fluctuations\,} $\langle (\sav-\sfS_n)^2 \rangle^{1/2}$
and $\langle(s_m-\sav)^2\rangle^{1/2}$ associated with the cell radius.


\subsection{The average $\sfL_n$ and approach to a horn torus}
\label{secaverageLn}
 
It follows from (\ref{fQnL}) that the average of $L$ for given $n$, to be
denoted as $\sfL_n\equiv\int_0^\infty\dd L\,LQ_n(L)$,
behaves asymptotically as
\bea
\sfL_n &=& \frac{1}{\kap n^{1/6}} \Big[
\overline{y} + \frac{1}{n} \Big( q_2\overline{y^3} 
+q_4\overline{y^5} - q_0 \overline{y} \Big) + {\cal O}(n^{-2})
\Big] \non
&=& \frac{1}{\kap n^{1/6}}\Big[ 1 + \frac{3a_2+2}{4n} + {\cal O}(n^{-2})
\Big) \Big],
\label{xLn}
\eea
where in passing from the first to the second line we
inserted the explicit expressions 
$\overline{y}=1,\, \overline{y^3}=\pi/2$ and $\overline{y^5}=3\pi^2/8$.
Using finally equations (\ref{xa2}) and (\ref{xf2f4})
we may render the coefficient $a_2$ more explicit and
(\ref{xLn}) becomes
\beq
\sfL_n = \frac{1}{\kap n^{1/6}}
\Big[ 1+ \frac{l_1}{n} + {\cal O}(n^{-2}) \Big]
\label{fLn}
\eeq
with $\kap$ given by (\ref{dy}) and in which 
\bea
l_1 = \frac{13}{18}+\frac{1}{8}\sum_{q=1}^\infty\,
\frac{18+11q^2}{9+q^4} \,=\,1.95977.
\label{xell1}
\eea
We now recall relation (\ref{dsm}),
\beq
s_m^2=r_m^2-L^2, \qquad m=1,2,\ldots,n
\label{dsmbis}
\eeq
(see Fig.\,\ref{figure_2}).
Since $s_m$ gets sharply peaked around $\sfS_n\sim n^{1/3}$
and $\sfL_n$ is of order $n^{-1/6}$,
we conclude that $r_m$ must be sharply peaked  around a value
that we will denote by $\sfR_n$ and that is equal to $\sfS_n$.
That is, because of (\ref{xsfSn}), 
\beq
\sfR_n \simeq \sfS_n \simeq (2\pi^2\lambda)^{-1/3}n^{1/3}.
\label{xsfRn}
\eeq
Since $\sfS_n$ and $\sfR_n$ are the major and minor radius, respectively, 
of the limit torus, their equality in the limit $n\to\infty$ implies that the
excluded domain tends to a horn torus, that is, a doughnut with a hole of zero
diameter. This conclusion was first reached by Hilhorst and Lazar
\cite{HilhorstLazar14} on the basis of a heuristic theory
and of simulations that extended initial work 
due to Lazar {\it et al.} \cite{Lazaretal13}.
We will now make a brief comparison with that work.


\subsection{Comparison to the ``entropy {\it vs.} entropy''  theory}
\label{seccomparisonHL}

\begin{figure}
\begin{center}
\scalebox{.40}
{\includegraphics{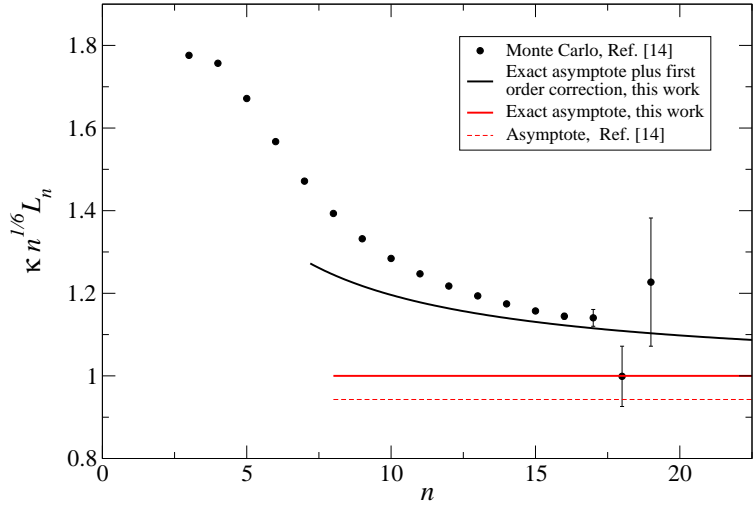}}
\end{center}
\caption{{\small 
Average focal distance $\sfL_n$ as a function of $n$.
Black dots: Monte Carlo data of Ref.\,\cite{HilhorstLazar14}.
Solid black curve: prediction of this work, including the 
${\cal O}(1/n)$ correction to the exact asymptotic behavior.
Solid red curve: exact asymptotic limit.
Dashed red curve: asymptotic limit according to the heuristic theory of
Ref.\,\cite{HilhorstLazar14}.
}}
\label{figure_4}
\end{figure}
\begin{figure}
\begin{center}
\scalebox{.40}
{\includegraphics{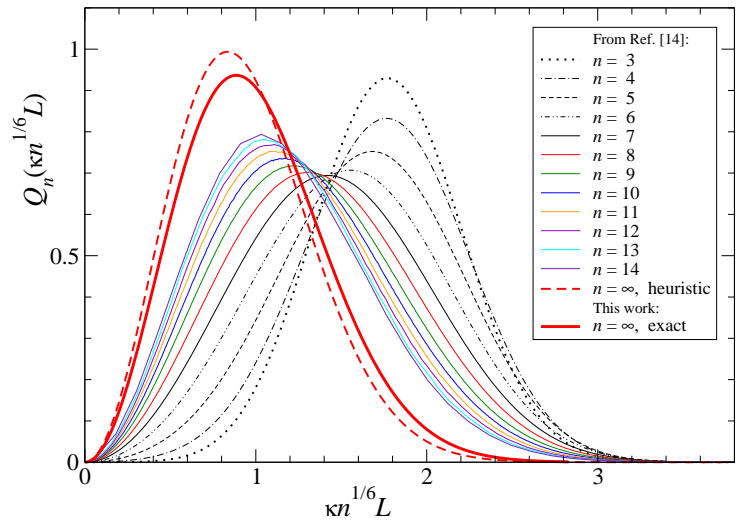}}
\end{center}
\caption{{\small 
Heavy red solid curve: limit distribution ${\cQ}(y)$ of
the scaled focal distance $y=\kappa n^{1/6}L$ as $n\to\infty$
[Eqs.\,(\ref{xQy}) and (\ref{dy})].
Heavy red dashed curve: prediction of the heuristic theory of 
Ref.\,\cite{HilhorstLazar14}. 
Other curves: Monte Carlo results \cite{HilhorstLazar14}
for the finite-$n$ distributions ${\cQ}_n(y)$ for $n=3,4,\ldots,14$.
}}
\label{figure_5}
\end{figure}
\begin{figure}
\begin{center}
\scalebox{.40}
{\includegraphics{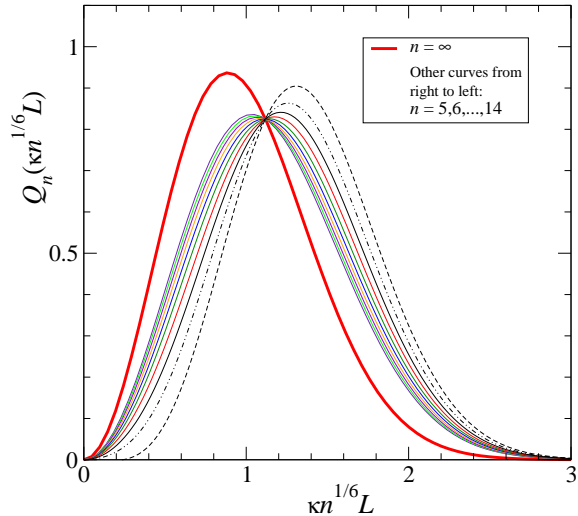}}
\end{center}
\caption{{\small 
Heavy red solid curve: same as in figure \ref{figure_5}.
Other curves: the distributions  ${\cQ}_n(y)$
according to Eq.\,(\ref{fQnL})
for $n=5,6,\ldots,14$ (same color code as in figure \ref{figure_5}).
See text for further comments.
}}
\label{figure_6}
\end{figure}

An alternative but heuristic approach to the study of 
various statistical properties of Voronoi cells
is based on an ``entropy {\it versus\,} entropy'' argument.
The heuristic theory was
initially applied to the $n$-sided 2D Poisson-Voronoi cell \cite{Hilhorst09a}, 
for which several of its results turned out to be exact,
in particular those for
the scaling of the mean cell radius with $n$.
This approach was then generalized \cite{Hilhorst09b} to the $3$-dimensional
$n$-faced Voronoi cell, which in the limit $n\to\infty$ becomes a sphere.

However, the {\it $n$-edged face\,}  between adjacent 3D cells 
has, in the large-$n$ limit, no spherical but merely axial symmetry.
For this reason the heuristic theory for the face \cite{HilhorstLazar14}, 
depends on an additional assumption.

The exact results found in this work
now allow us to assess the validity of the heuristic theory.

${}$\phantom{ii}(i) The exact asymptotic $n$ dependence of $\sfR_n$ and $\sfS_n$
[Eq.\,(\ref{xsfRn})] coincides with the results
of the heuristic theory of Ref.\,\cite{HilhorstLazar14}.

${}$\phantom{i}(ii) The exact asymptotic $n$ dependence of $\sfL_n$ 
[Eq.\,{\ref{fLn}}] has the same power $-1/6$
as found heuristically; however, the exact prefactor 
$\kap^{-1} = 0.511357\lambda^{-1/3}$  [Eq.\,(\ref{dy})] 
is larger than the heuristic one
by a factor of $\sqrt{9/8}=1.06$.

(iii) The exact function ${\cQ}(y)$ that describes the asymptotic
probability distribution of $L$ [Eq.\,(\ref{xQy})] is the same as 
the heuristic one 
up to a scaling with the same factor $\sqrt{9/8}$
(see figure \ref{figure_5}).
It is remarkable, since this was unforseeable, 
that the heuristic theory for the distribution of $y$ should be so close to
being exact. 

Our present results go beyond those predicted by the heuristic theory, in particular in that they
provide, in Eqs.\,(\ref{fQnL}) and (\ref{fLn}), the leading order
finite-$n$ corrections to the distribution function ${\cQ}(y)$ and to the
average $\sfL_n$, respectively. In the next subsection we will compare these 
new results to earlier Monte Carlo simulations.


\subsection{Comparison to Monte Carlo work}
\label{seccomparisonMC}

We consider the average focal distance $\sfL_n$ and refer to Figure \ref{figure_4}.
The Monte Carlo data \cite{HilhorstLazar14} for this quantity 
are accurate up to about $n\lesssim 17$.
They show appreciable finite-$n$ deviations from the heuristically predicted
asymptotic large-$n$ behavior (the dashed red line).
This work brings theory and simulations much closer together.
First of all, the exact asymptote (solid red line) is higher than the
heuristic one by the factor $\sqrt{9/8}$ discussed above.
Furthermore, inclusion of the ${\cal O}(1/n)$ correction
term [see Eqs.\,(\ref{fLn})-(\ref{xell1})] greatly improves the
correspondence between theory and simulations.

We now turn to the probability distributions $Q_n(L)$ themselves,
or rather theor scaled equivalents ${\cQ}_n(\kappa n^{1/5}L)$.
In Fig.\,\ref{figure_5} Monte Carlo data \cite{HilhorstLazar14} are shown
for $\cQ_n$ with $n=3,4,\ldots,14$. The exact limiting curve for $n\to\infty$
[Eq.(\ref{xQy})] is the solid black line. 
The heuristic theory predicted the dashed red curve.
Although the exact limit is closer to the finite-$n$ Monte Carlo data, there
are still considerable finite size effects for the valuse of $n$ attainable by
the simulations. 

In Fig.\,\ref{figure_6} we show the distributions ${cQ}_n(y)$ for 
$n=5,6,\ldots,14$ based on Eq.\,(\ref{fQnL}), 
and incorporationg the correction term of order $n^{-1}$.
The agreement with the Monte Carlo data is qualitative:
as $n$ increases, the average $\sfL_n$ goes down while the distribution first
gets wider and then narrower again,
which has the consequence that the peak value passes
through a minimum. 
Quantitative agreement gets better as $n$ gets large, but finite size effect
remain clearly visible.


\section{Conclusion}
\label{secconclusion}

This work represents a new contribution to the statistics of Poisson-Voronoi
tessellations in three dimensions.
\vspace{2mm}

We have studied an arbitrary face shared by two
neighboring cells, the ``focal cells.'' 
We determined the probability $p_n$ for this face
to have exactly $n$ edges, as well as the
conditional probability distribution $Q_n(L)$ of the focal distance $L$ 
({\it i.e.} half the distance between the seeds of the focal cells) 
given the edgedness $n$. Calculating these quantities amounts to solving 
a problem of $n$ interacting particles, and we have shown that this problem
may be brought under full control in the limit $n\to\infty$. 
\vspace{2mm}

The analytic methods of this paper were developed initially within the context of several two-dimensional problems in random geometry.
We have extended them here for the first time to a problem in three dimensions.
\vspace{2mm}

Our results, summarized at the end of the introduction,
include expressions for the asymptotic $n\to\infty$ behavior of $p_n$ and
$Q_n(L)$. The focal
distance $L$ was shown to scale as $n^{-1/6}$ with
corrections of relative order $1/n$ whose amplitude was determined. 
The agreement between the present theory and
earlier Monte Carlo simulations is good.

The positions of the edges of the $n$-edged face are determined by the
positions of $n$ first-neighbors seeds to the pair of focal seeds.
These first neighbors were
shown to lie, for large $n$, on the surface of a spindle torus whose interior
excludes all seeds other than the two focal seeds. 
For $n\to\infty$ the major and minor radii of this torus 
(that were shown to scale as $n^{1/3}$) become equal:
the limit of the excluded domain is a doughnut with a zero diameter hole.

We conclude by mentioning again
the closely related problem that comes naturally to mind, {\it viz.,}
to find, for asymptotically large $n$, the probability $p^{(3)}_n$ that a three-dimensional Poisson-Voronoi cell have $n$ faces. 
In spite of the progress achieved here,
that question remains an open challenge.



\end{small}
\end{document}